\documentclass[twocolumn]{aastex631}

\shorttitle{A kinematic census of the local Universe}
\shortauthors{Fraser-McKelvie \& Cortese}
\graphicspath{{./}{figures/}}
\begin{document}
\title{Beyond galaxy bimodality: the complex interplay between kinematic morphology and star formation in the local Universe}

\author[0000-0001-9557-5648]{A. Fraser-McKelvie}
\affiliation{International Centre for Radio Astronomy Research, The University of Western Australia, 35 Stirling Hwy, 6009 Crawley, WA, Australia}
\affiliation{ARC Centre of Excellence for All Sky Astrophysics in 3 Dimensions (ASTRO 3D)}

\author[0000-0002-7422-9823]{L. Cortese}
\affiliation{International Centre for Radio Astronomy Research, The University of Western Australia, 35 Stirling Hwy, 6009 Crawley, WA, Australia}
\affiliation{ARC Centre of Excellence for All Sky Astrophysics in 3 Dimensions (ASTRO 3D)}

%% Note that the \and command from previous versions of AASTeX is now
%% depreciated in this version as it is no longer necessary. AASTeX 
%% automatically takes care of all commas and "and"s between authors names.

%% AASTeX 6.31 has the new \collaboration and \nocollaboration commands to
%% provide the collaboration status of a group of authors. These commands 
%% can be used either before or after the list of corresponding authors. The
%% argument for \collaboration is the collaboration identifier. Authors are
%% encouraged to surround collaboration identifiers with ()s. The 
%% \nocollaboration command takes no argument and exists to indicate that
%% the nearby authors are not part of surrounding collaborations.

%% Mark off the abstract in the ``abstract'' environment. 
\begin{abstract}
It is generally assumed that galaxies are a bimodal population in both star formation and structure: star-forming galaxies are disks, while passive galaxies host large bulges or are entirely spheroidal. Here, we test this scenario by presenting a full census of the kinematic morphologies of a volume-limited sample of galaxies in the local Universe extracted from the MaNGA galaxy survey. We measure the integrated stellar line-of-sight velocity to velocity dispersion ratio ($V/\sigma$) for 4574 galaxies in the stellar mass range $9.75 < \log M_{\star}[\rm{M}_{\odot}] < 11.75$. We show that at fixed stellar mass, the distribution of $V/\sigma$ is not bimodal, and that a simple separation between fast and slow rotators is over-simplistic. Fast rotators are a mixture of at least two populations, referred to here as dynamically-cold disks and intermediate systems, with disks dominating in both total stellar mass and number. When considering star-forming and passive galaxies separately, the star-forming population is almost entirely made up of disks, while the passive population is mixed, implying an array of quenching mechanisms. Passive disks represent $\sim$30\% (both in number and mass) of passive galaxies, nearly a factor of two higher than that of slow rotators, reiterating that these are an important population for understanding galaxy quenching.
These results paint a picture of a local Universe dominated by disky galaxies, most of which become somewhat less rotation-supported upon or after quenching. While spheroids are present to a degree, they are certainly not the evolutionary end-point for the majority of galaxies.

\end{abstract}

%% Keywords should appear after the \end{abstract} command. 
%% The AAS Journals now uses Unified Astronomy Thesaurus concepts:
%% https://astrothesaurus.org
%% You will be asked to selected these concepts during the submission process
%% but this old "keyword" functionality is maintained in case authors want
%% to include these concepts in their preprints.

%% From the front matter, we move on to the body of the paper.
%% Sections are demarcated by \section and \subsection, respectively.
%% Observe the use of the LaTeX \label
%% command after the \subsection to give a symbolic KEY to the
%% subsection for cross-referencing in a \ref command.
%% You can use LaTeX's \ref and \label commands to keep track of
%% cross-references to sections, equations, tables, and figures.
%% That way, if you change the order of any elements, LaTeX will
%% automatically renumber them.
%%
%% We recommend that authors also use the natbib \citep
%% and \citet commands to identify citations.  The citations are
%% tied to the reference list via symbolic KEYs. The KEY corresponds
%% to the KEY in the \bibitem in the reference list below. 

\section{Introduction} \label{sec:intro}
The exquisite morphological diversity of the local Universe galaxy population advocates for a range of formation and evolution pathways. 
While it is generally accepted that dynamically-cold disk-like structures arise from gas accretion \citep{fall1980}, 
dispersion-rich structures are either formed via dissipationless galaxy mergers \citep{walker1996} or the coalescence of giant clumps in high-redshift galaxies \citep{noguchi1999,bournaud2007}. The prevalence of dispersion-dominated structures such as spheroidal galaxies and classical bulges in the local Universe therefore provide important constraints on the physical processes behind the growth of galaxies.

Traditionally, galaxy morphological classification relied on either visual classification or two-dimensional bulge-disk decompositions, resulting in a high fraction of the overall stellar mass in the local Universe being attributed to spheroids and classical bulges \citep[e.g.][]{renzini2006, driver2007, kelvin2014}. The consequence of a local Universe dominated by dispersion-supported structure is a prominent role for mergers in galaxy growth \citep[e.g.][]{vandokkum2010, lopez-sanjuan2012}. Following on from this scenario is the idea that there exists a strong correlation between a galaxy's visual morphology and its current star formation rate (SFR): star-forming galaxies are disks, and passive galaxies possess prominent classical bulges or a purely spheroidal morphology \citep[e.g.][]{wuyts2011, lang2014}. Given that the distribution in SFR is generally considered to be bimodal \citep[e.g.][]{brinchmann2004}, the general assumption is that structure and SFR map on to each other, such that the distribution of galactic morphology is also assumed to be bimodal.

There are several issues with the picture presented above. One is the inability of visual classification and photometric decomposition  to differentiate between rotation-supported disks at low inclinations, and prolate, dispersion-supported spheroids. Additionally, the high values of S\'{e}rsic index used to imply classical bulge presence may be the result of several underlying disky structural components (e.g. nuclear disks, rings, inner bars) adding to the central light concentration to give the illusion of bulge-like structure \citep[e.g.][]{erwin2021}. Furthermore, the presence of entirely passive spiral galaxies is evidence that galaxies can quench without morphological transformation into largely spheroid-dominated systems \citep{vandenbergh1976,masters2010, fraser-mckelvie2016,pak2019}. 
It would seem that not all passive galaxies must possess dispersion-supported structure, and not all disky galaxies are actively star-forming. It is possible that the local Universe is not as dominated by spheroidal structure as once thought.

In fact, our knowledge of galaxy evolution has transformed over recent years to the point where it is clear that the quenching of a galaxy's star formation is not always coincident with a morphological transformation \citep[e.g.][]{wang2020,fraser-mckelvie2021,cortese2022}. 
The advent of large-scale galaxy integral field spectroscopy (IFS) surveys has allowed the field to break the degeneracies of visual classification or 2D decomposition through the quantification of the internal motions of a galaxy's stellar component. Classified according to the degree of rotation or dispersion support, the stellar kinematics provide a more physical framework to define the structure of galaxies.

IFS surveys such as the SAURON project \citep{bacon2001} pioneered the use of stellar kinematics to categorize galaxies, removing some of the main uncertainties related to visual morphologies. The degree of stellar rotation to dispersion support in a galaxy \citep[often quantified by the ratio of rotational velocity to velocity dispersion $V/\sigma$, or the spin parameter, $\lambda_{Re}$ e.g.][]{cappellari2007}, are key tools in understanding the motions of the stars and the angular momentum content of a galaxy.

Early IFS results separated galaxies into `fast rotators' (those whose rotation support dominated over dispersion support) and `slow rotators'  \citep[dispersion-supported systems e.g.][]{cappellari2007, emsellem2007}. Fast rotators are often considered to be one population, though span a range of spin parameter values. Stellar rotational support can decrease via mergers and tidal effects, resulting in dispersion-supported structures including classical bulges and thick disks \citep[e.g.][]{choi2018,lagos2018,walo-martin2020}, or via early star formation quenching \citep[e.g.][]{lagos2017}.
For this reason, the degree of rotation-to-dispersion support within a galaxy can provide important clues on its internal structure, assembly, and star formation history.

While it seems clear that slow-rotators contribute little to the global mass and number budget of the local Universe \citep[e.g.][]{emsellem2011,vandesande2017a, guo2020}, the spread of galaxy spin values within the canonical $\lambda_{Re}$-ellipticity plane and implications for galaxy assembly histories for the fast rotator population are still unclear. Quantifying this spread becomes particularly important when examining morphological distributions in the context of the star formation properties of galaxies.

To make progress, we need to quantify the spread of structure using a physically-motivated kinematic parameter and map it to the SFR for a sample whose stellar mass distribution is consistent with the local Universe. While the link between stellar spin and SFR has already been investigated by several studies \citep[e.g.][]{shapiro2010,cortese16,vandesande2018,guo2020,wang2020, fraser-mckelvie2021,cortese2022}, no study has yet been able to quantify the contributions of different kinematic populations to the number and stellar mass budget of galaxies for a representative sample of the local Universe. Whilst \citet{guo2020} attempted such an effort using the SAMI galaxy survey, the high stellar masses were not sufficiently covered: an IFS galaxy survey with more complete number statistics at the high stellar mass end is required.

The Mapping Nearby Galaxies at APO \citep[MaNGA;][]{bundy2015} galaxy survey is the largest optical IFS galaxy survey in existence, having observed over 10,000 galaxies over the stellar mass range $8 \lesssim \log M_{\star}[\rm{M}_{\odot}]\lesssim 12$. Harnessing the statistical power of this dataset, we aim to complete a kinematic census of galaxies in the local Universe by determining the fraction of fast and slow rotators (allowing for multiple populations within the fast rotator category) and link these to the current star formation properties of the galaxies. In Section~\ref{Sect2}, we describe the MaNGA galaxy survey, the determination of kinematic parameters $\lambda_{Re}$ and $V/\sigma$, and the volume weighting carried out to create a complete sample, in Section~\ref{Sect3} we present our results and discussion, and in Section~\ref{Sect4} summarize and present our conclusions. Throughout this paper we use $\Lambda$CDM cosmology, with $\Omega_{m}=0.27$, $\Omega_{\lambda}=0.73$, $H_{0}=70.4~\rm{km}~\rm{s}^{-1}~\rm{Mpc}^{-1}$ and a \citet{chabrier2003} initial mass function.

%%%%%%%%%%%%%%%%%%%%%%%%%%%%%%%%%%%%%%%%%%%%
%% DATA & METHODS %%
%%%%%%%%%%%%%%%%%%%%%%%%%%%%%%%%%%%%%%%%%%%
\section{Data \& Methods}
\label{Sect2}
\subsection{The MaNGA galaxy survey}
The MaNGA Galaxy Survey is an integral field spectroscopic survey that observed $\sim10,000$ galaxies \citep{bundy2015, drory2015, abdurrouf2022}. It is an SDSS-IV project \citep{blanton2017}, employing the 2.5m telescope at Apache Point Observatory \citep{gunn2006} and BOSS spectrographs \citet{smee2013}. SDSS data release 17 \citep{abdurrouf2022} contains the final MaNGA data release of 10,010 unique galaxy observations, observed and reduced by the MaNGA data reduction pipeline \citep[DRP;][]{law2015}. Derived properties including emission line fits \citep{belfiore2018} were produced by the MaNGA data analysis pipeline \citep[DAP;][]{westfall2019}, and provided as a single data cube per galaxy \citep{yan2016a}. MaNGA’s target galaxies were chosen to include a wide range of galaxy masses and colors, over the redshift range $0.01 < z < 0.15$, and the Primary+ sample \citep{yan2016, wake2017} contains spatial coverage out to $\sim$ 1.5 effective radii ($R_{e}$) for $\sim$66\% of all observed galaxies, the remainder of which constitute the Secondary sample, which are observed out to $\sim2.5~R_{e}$. For this work, we utilize the 2D kinematic map data products produced by the MaNGA DAP, including the stellar rotational velocity and velocity dispersion. The maps are Voronoi binned to a signal-to-noise ratio of 10.

\subsection{Kinematic sample}
As we wish to compare the stellar kinematic properties of MaNGA galaxies with their current SFR, we select galaxies from the full MaNGA sample with both reliable stellar kinematic measures and available homogeneous SFR measures. 
Starting from the full MaNGA sample of 10,010 galaxies, we match to the \textit{GALEX}-Sloan-\textit{WISE} Legacy Catalogue 2 \citep[GSWLC-2;][]{salim2016,salim2018}, using a sky match with maximum separation of 2$^{\prime\prime}$. We utilize the GSWLC-X2 catalogue, which uses the deepest \textit{GALEX} photometry available for the spectral energy distribution (SED)-derived SFRs and stellar masses. 8637 galaxies have SFR and stellar mass measures in GSWLC-2, though 25 of these have no available $R_{e}$ or axis ratio measures in MaNGA's DRP summary table (\texttt{drpall\_v3\_1\_1}), compiled from the NASA Sloan Atlas, leaving 8612 galaxies.
Given that we wish to correct our sample to a complete, volume-limited sample, we remove galaxies without a volume weight measure in the DRP summary table. We also remove those with data reduction quality bits that indicate severe quality control issues (flag \texttt{drp3qual=DONOTUSE}), leaving 8312 galaxies. These 8312 galaxies constitute the sample for which we attempt to measure stellar kinematic properties.

\subsection{Determining $V/\sigma$ and $\lambda_{Re}$}
We determine $V/\sigma$ within 1$R_{e}$ using the definition of \citet{cappellari2007}:
\begin{equation}
    \left(\frac{V}{\sigma}\right)^{2} \equiv \frac{\langle V^{2}\rangle}{\langle\sigma^{2}\rangle} = \frac{\sum_{i=0}^{N_{spx}} F_{i}V_{i}^{2}}{\sum_{i=0}^{N_{spx}} F_{i}\sigma_{i}^{2}},
\end{equation}
where $F$ is the mean flux in the $r$-band for the binned spectra (\texttt{BIN\_MFLUX} map from the MaNGA DAP), $V$ the stellar rotational velocity (\texttt{STELLAR\_VEL} map), and $\sigma$ the stellar velocity dispersion (\texttt{STELLAR\_SIGMA} map) of the $ith$ spaxel.

We also estimate the stellar spin parameter within $1R_{e}$ as defined by \citet{emsellem2007, emsellem2011}:
\begin{equation}
    \lambda_{R} = \frac{\langle R|V|\rangle}{\langle R\sqrt{V^{2} + \sigma^{2}}\rangle}  = \frac{\sum_{i=0}^{N_{spx}} F_{i}R_{i}|V_{i}|}{\sum_{i=0}^{N_{spx}} F_{i}R_{i}\sqrt{V_{i}^{2} + \sigma_{i}^{2}}},
\end{equation}
where $F$, $V$, and $\sigma$ are as defined above. In the same manner as \citet{cortese16} and \citet{fraser-mckelvie2021}, we define $R$ as the semi-major axis of an ellipse on which spaxel $i$ lies.  

\begin{figure}
\includegraphics[width=0.48\textwidth]{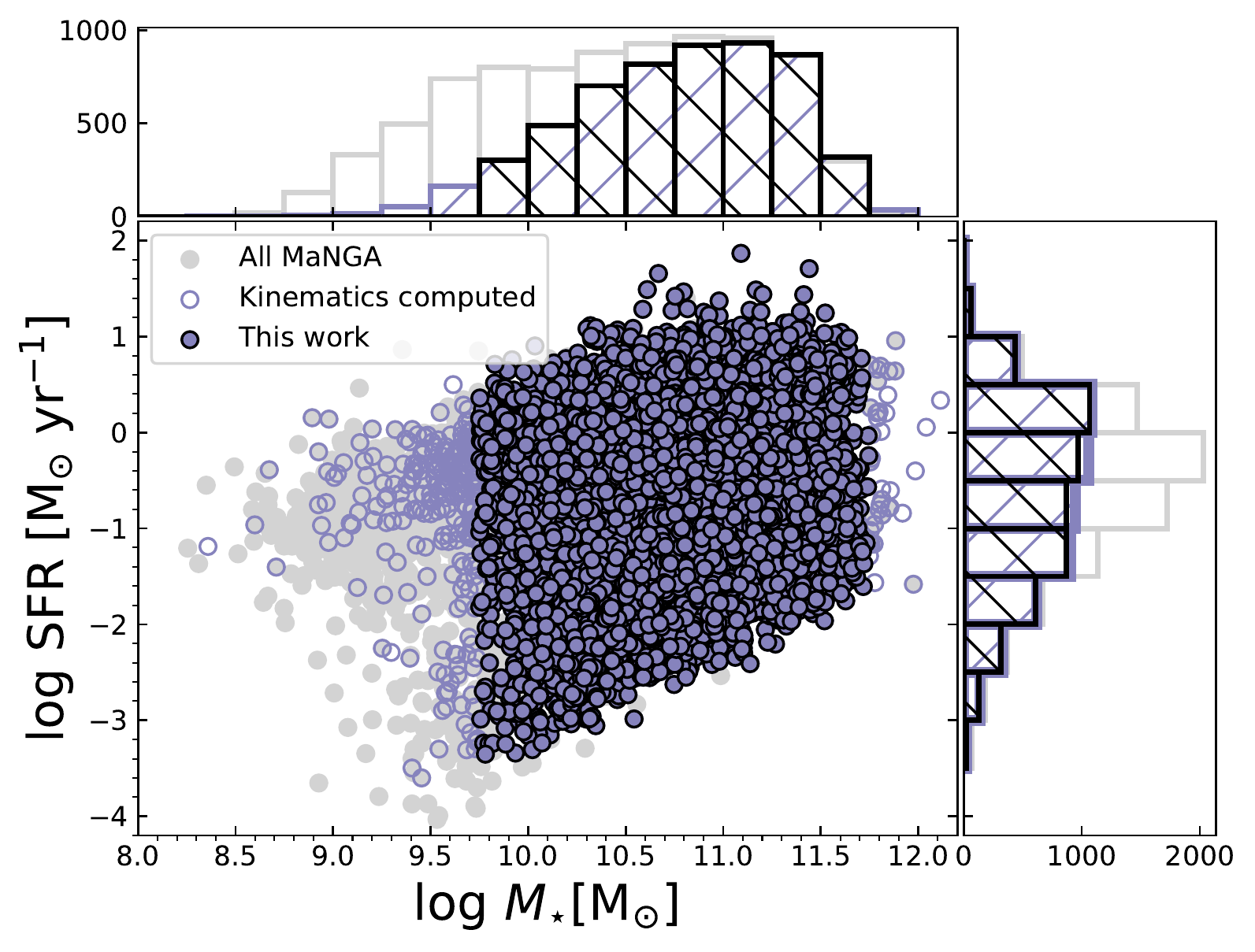} 
%\plotone{fig1c.pdf} %/Documents/2022/MaNGA_vsig/Fig1a.py
\caption{Sample selection for this work on the SFR-stellar mass plane. The overall MaNGA sample is shown in grey, galaxies for which kinematics were computed are shown as unfilled mauve circles, and the final sample after quality and stellar mass cuts are shown as filled mauve circles with black edges.
Histograms of all three distributions are shown in stellar mass (top) and SFR (right).
 There is a strong mass dependence such that low-mass galaxies frequently do not possess kinematic measures. For this reason, coupled with low number statistics at the high-mass end, we restrict our proceeding analysis to $9.75 < \log M_{\star}[\rm{M}_{\odot}]<11.75$.   \label{fig:sample1}}
\end{figure}

For both $V/\sigma$ and $\lambda_{Re}$ measurements, we mask spaxels where the stellar velocity dispersion, $\sigma_{\star}$, is less than $50~\rm{km}~\rm{s}^{-1}$, as the MaNGA line spread function is not well understood below these levels \citep{westfall2019}. Any galaxy with $>20\%$ masked spaxels within $1R_{e}$ is not included in our analysis. Unsurprisingly, given the lower $\sigma_{\star}$ values expected in low-mass galaxies coupled with the lower S/N for fainter galaxies, the majority of galaxies with $\log M_{\star}[\rm{M}_{\odot}]<9.75$ were not included in this analysis, and we limit the mass range of this work to $9.75 < \log M_{\star}[\rm{M}_{\odot}]<11.75$ to also account for low number statistics at the higher mass end. In addition, we remove galaxies from the sample for which $1 R_{e}<2.5^{\prime\prime}$, which is the approximate FWHM of the MaNGA spatial PSF \citep{law2015}. Fig~\ref{fig:sample1} shows the final kinematic sample used in this work compared to the overall MaNGA sample.
Of the 8312 galaxies with SFRs, stellar mass estimates, and calculated volume weights, 4574 had both $V/\sigma$ and $\lambda_{Re}$ measures and satisfy the above criteria and this is the final sample used for the following analysis. 

Atmospheric seeing conditions have been demonstrated to affect galaxy kinematic measurements \citep[e.g.][]{graham2018}, with the line-of-sight velocity dispersion artificially increased due to beam-smearing effects. We correct for beam smearing using the empirical relations of \citet{harborne2020}, created via mock observations of N-body galaxy models. In addition, galaxy line-of-sight rotational velocity measures are affected by inclination, and so both $V/\sigma$ and $\lambda_{Re}$ values are deprojected using the recipes provided in Appendix B of \citet{emsellem2011}.

\subsection{Volume weighting}
The MaNGA sample is not luminosity-independent and volume-limited within its redshift range; rather, it is over-sampled at high stellar masses. To correct for the survey selection function, as part of the MaNGA data release, volume weights are provided to up- or down-weight an individual galaxy based on the probability that a given galaxy would be allocated an IFU at a given redshift \citep[see][for further details]{wake2017}. These weights are for the entire sample as a whole, and cannot be used when employing only a portion of the sample, as in this work. Instead, we calculate revised volume weights based on a `completeness' value, defined as the fraction of galaxies for which we have calculated kinematic parameters for compared to the total number of galaxies in the overall MaNGA sample for a given bin in stellar mass and redshift.

As noted in \citet{sanchez2019}, the fraction of galaxies observed by MaNGA for a given stellar mass and redshift bin can vary with galaxy color or SFR. For this reason, we calculate the completeness for the star-forming population separately from the passive population, dividing the two samples by a line 0.5 dex below the star-forming main sequence defined in \citet{fraser-mckelvie2021}.
For each individual galaxy we then divide the provided MaNGA volume weight (we used the \texttt{esweight} parameter, as we are using the combined MaNGA Primary+ and Secondary samples) by the completeness of the redshift and stellar mass bin the galaxy is located in to obtain the revised volume weight for our sample. In this manner, we can reliably recover the MaNGA volume-weighted stellar mass function above stellar masses of $\log M_{\star}[\rm{M}_{\odot}]>9.75$, which is representative of the local Universe. Further details are provided in Appendix~\ref{appendix1}.  

%%%%%%%%%%%%%%%%%%%%%%%%%%%%%%%%%%%%%%%%%%%%
%% RESULTS & DISCUSSION                   %%
%%%%%%%%%%%%%%%%%%%%%%%%%%%%%%%%%%%%%%%%%%%%

\begin{figure*} %/Documents/2022/MaNGA_vsig/GMMs_sSFR2.py
\gridline{\fig{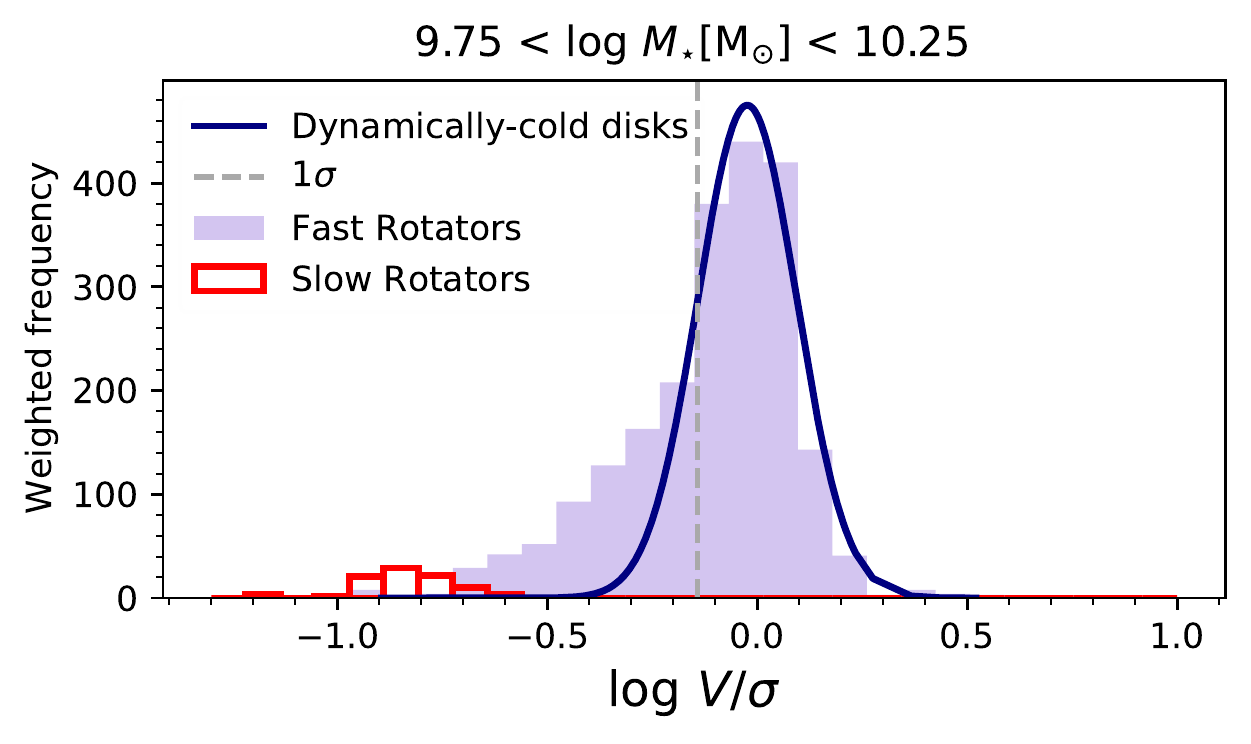}{0.47\textwidth}{(a)}
        \fig{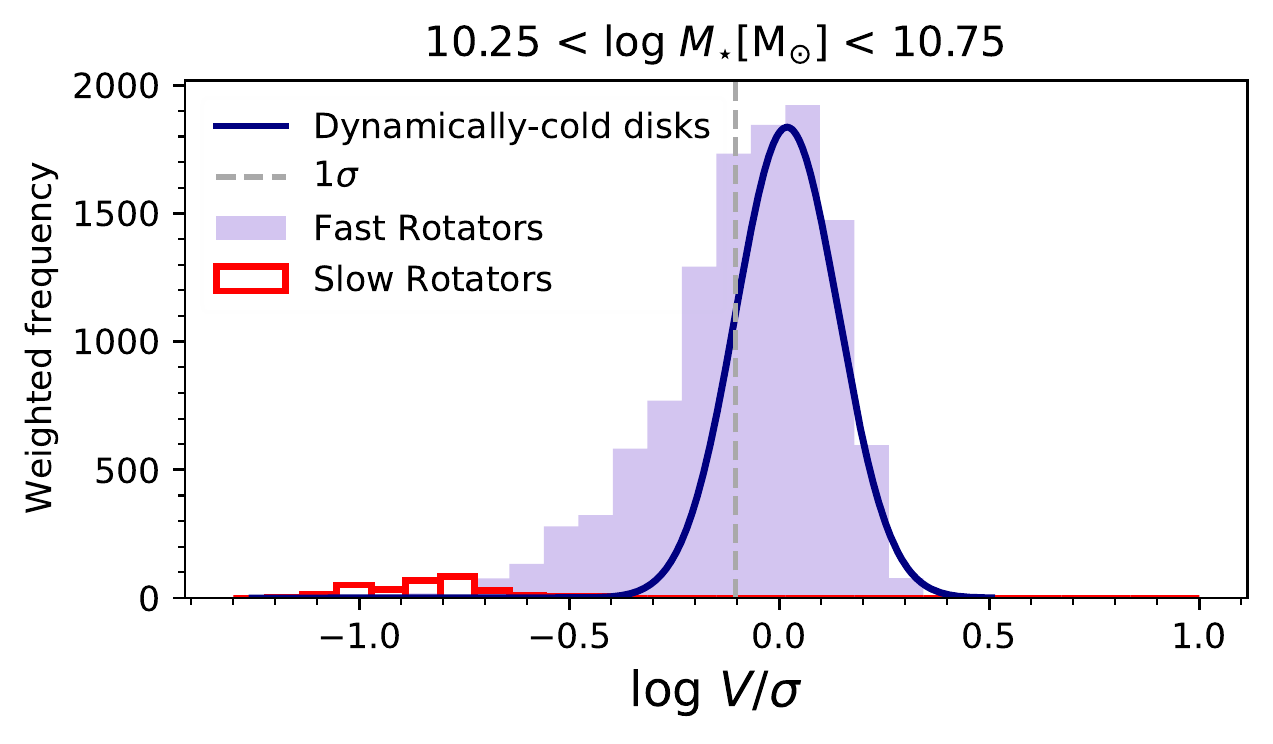}{0.47\textwidth}{(b)}
        }
\gridline{\fig{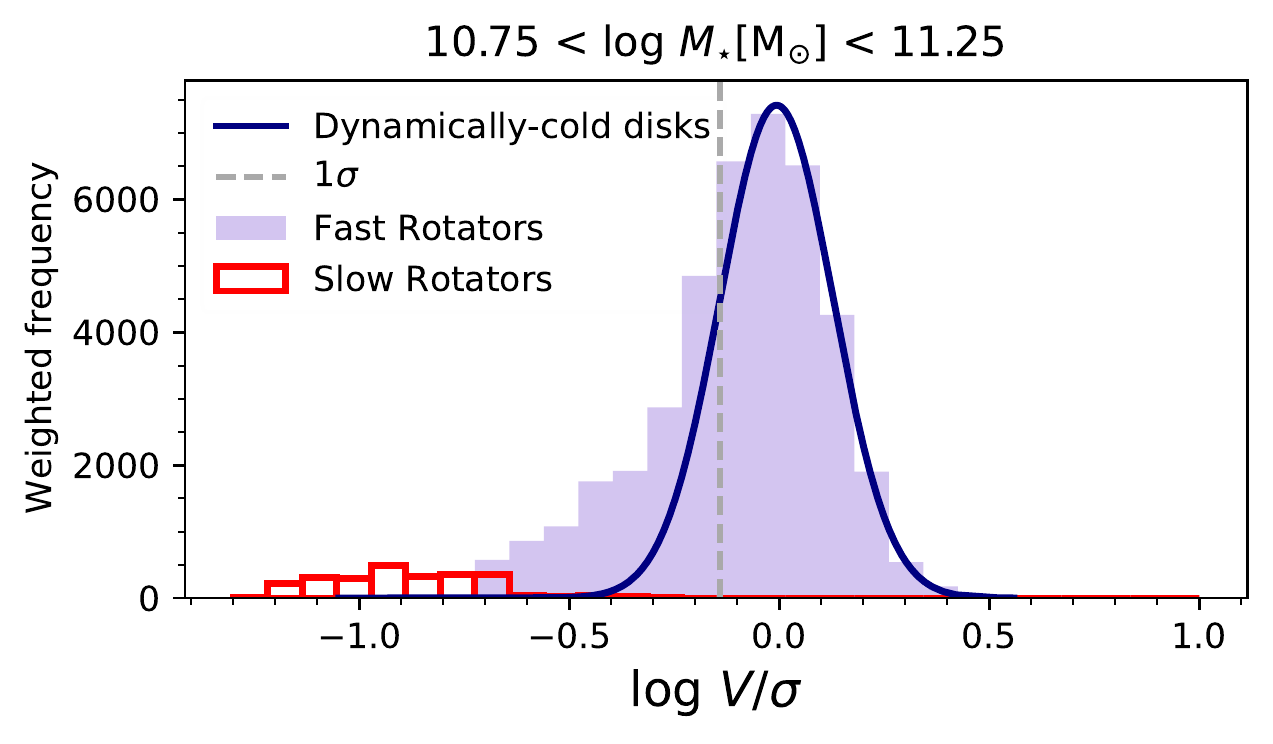}{0.47\textwidth}{(c)}
        \fig{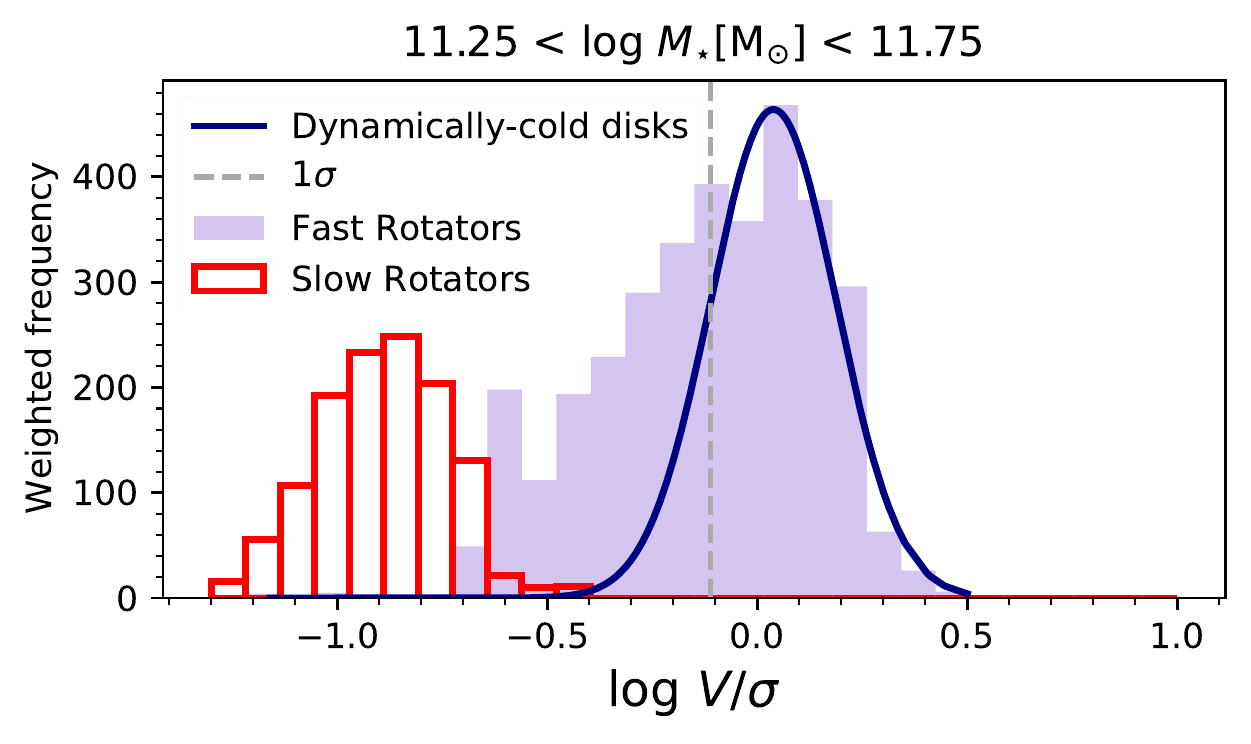}{0.47\textwidth}{(d)}
          }
\caption{Histograms of $\log V/\sigma$ for the volume-corrected MaNGA sample in four stellar mass bins. For each panel, the fast rotators are denoted by a purple histogram, and the slow rotators as an unfilled, red histogram. The fast rotators are modelled with a Gaussian mixture model, which is best described by two populations: the dynamically-cold disk population of which is shown as a navy line, and 1$\sigma$ below the mean as a dashed grey line. The grey dashed line denotes the division between dynamically-cold disks and intermediate systems for this work.
  \label{fig:results1}}
\end{figure*}

\begin{figure*} %/Documents/2022/MaNGA_vsig/GMMs_sSFR2.py
\gridline{\fig{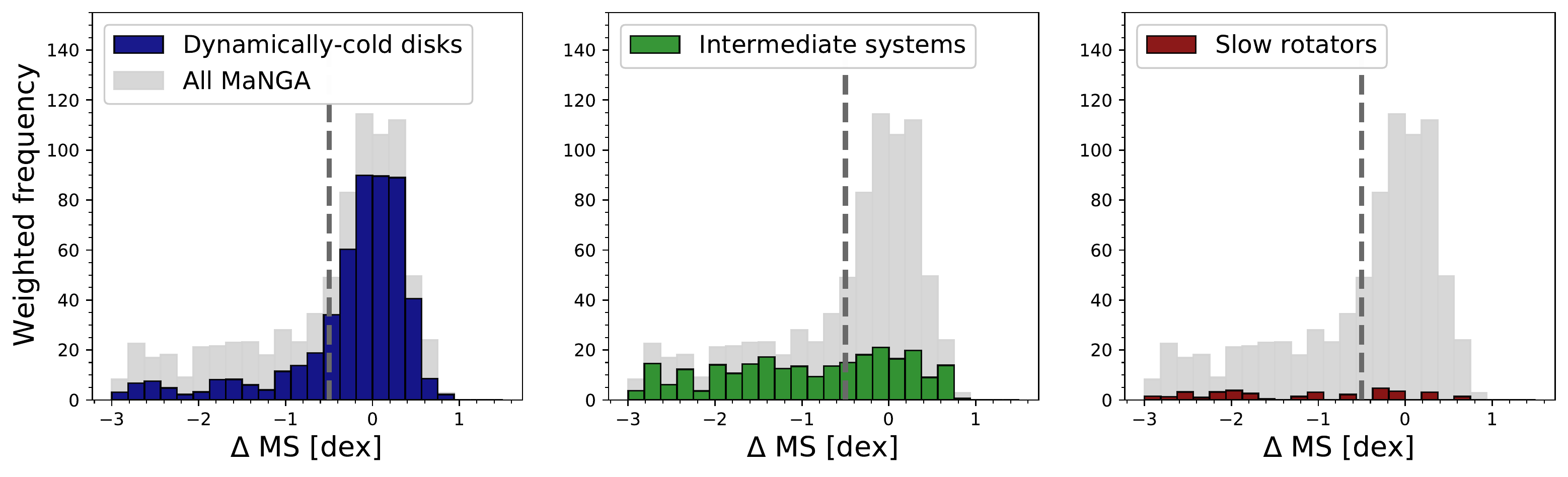}{0.8\textwidth}{(a) $9.75 < \log M_{\star}[\rm{M}_{\odot}]<10.25$}}
\gridline{\fig{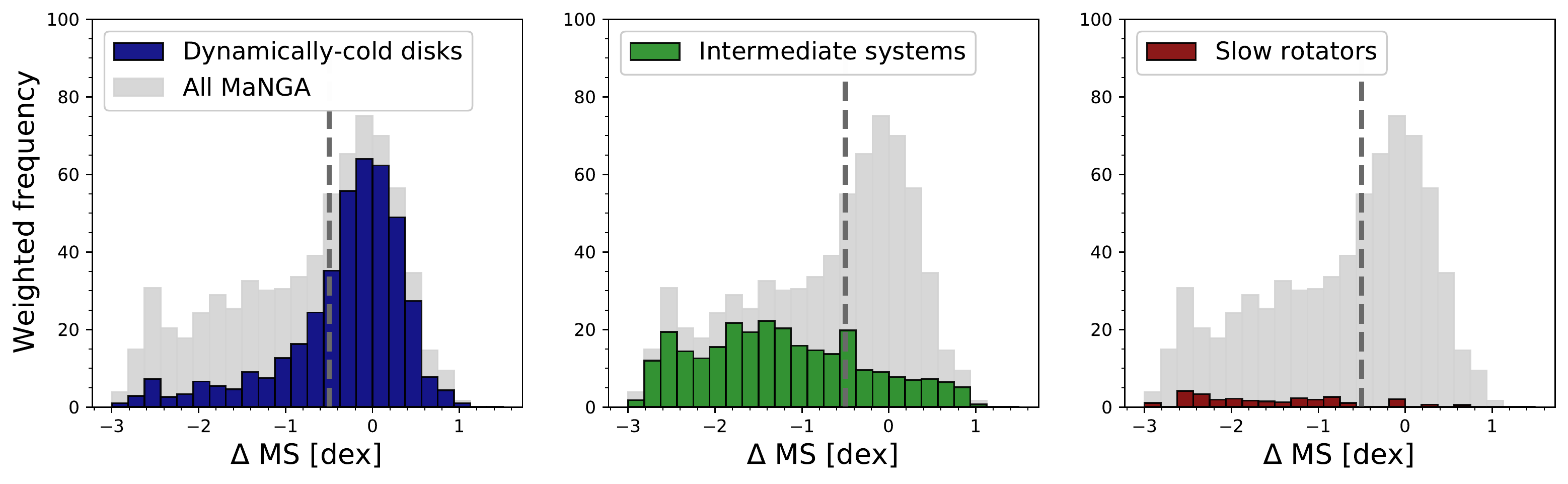}{0.8\textwidth}{(b) $10.25 < \log M_{\star}[\rm{M}_{\odot}]<10.75$}
          }
\gridline{\fig{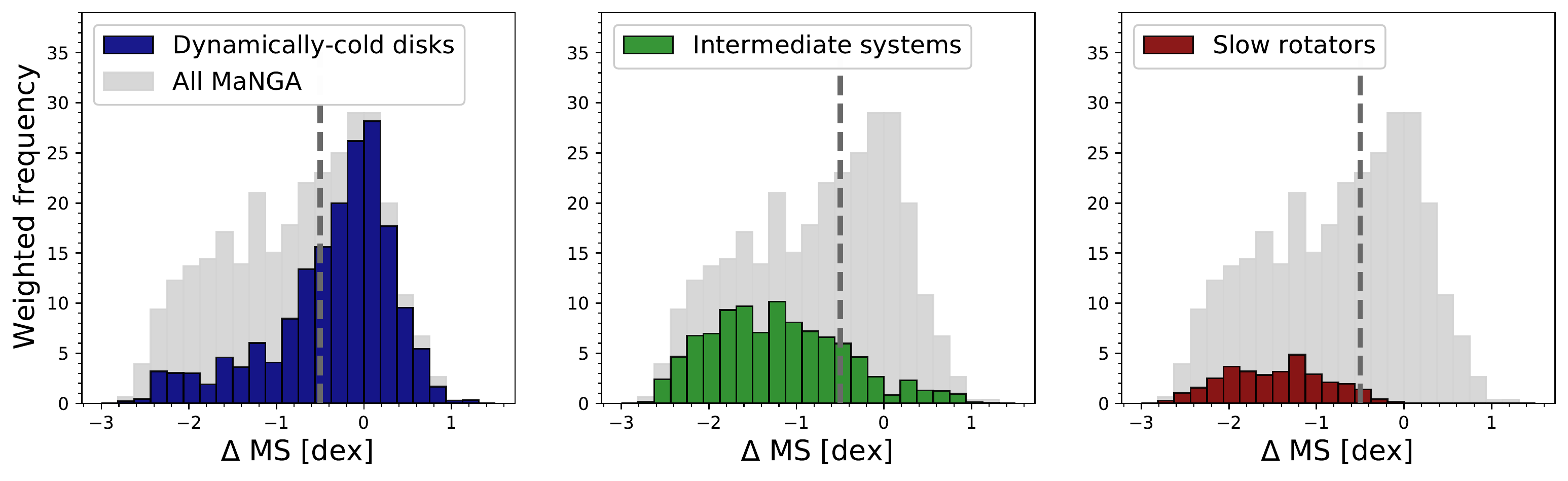}{0.8\textwidth}{(c) $10.75 < \log M_{\star}[\rm{M}_{\odot}]<11.25$}
          }
\gridline{\fig{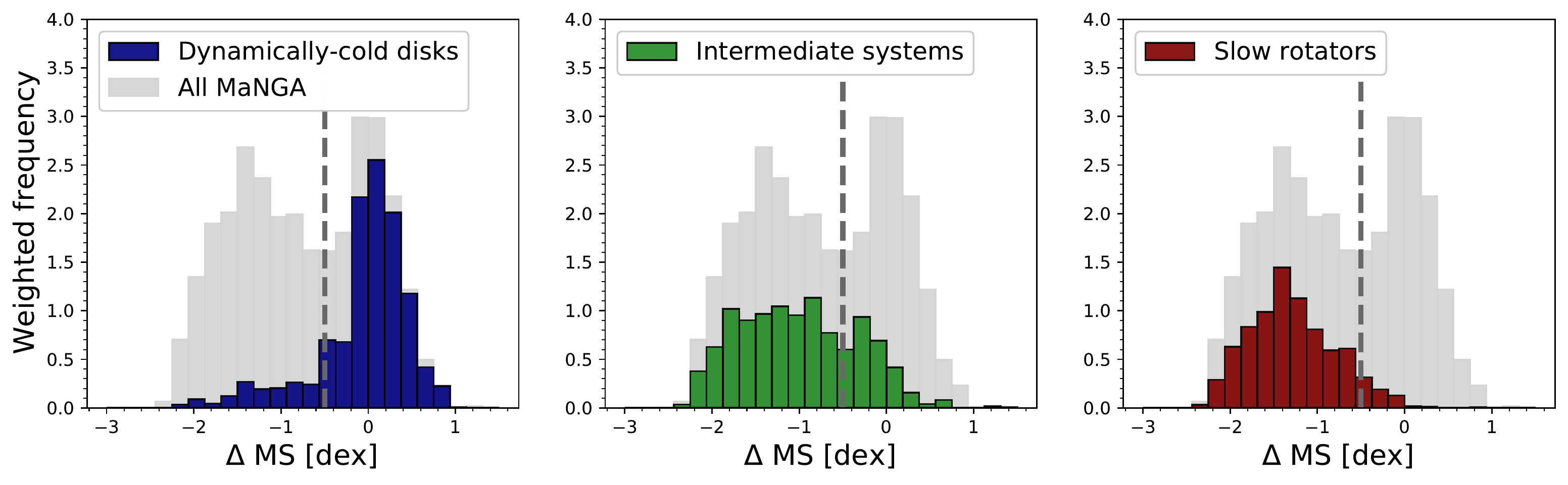}{0.8\textwidth}{(d) $11.25 < \log M_{\star}[\rm{M}_{\odot}]<11.75$}}

\caption{Histograms of dynamically-cold disks (left), intermediate systems (centre) and slow rotators (right) in four bins of stellar mass as a function of distance from the SFMS ($\Delta$MS). The dividing line between passive and star-forming galaxies used in this work is denoted at $\Delta$MS=-0.5 as a grey dashed line. The overall distribution of MaNGA galaxies for a given stellar mass bin is shown in grey.
\label{fig:pyramid}}
\end{figure*}

\section{Results \& discussion}
\label{Sect3}
With $V/\sigma$ and $\lambda_{Re}$ calculated for all galaxies in the kinematic sample, we can now carry out a full census of kinematic morphologies in the local Universe and link these to current SFR. 
The literature presents strong observational evidence for at least two kinematic classes of galaxies \citep{emsellem2007, cappellari2016, graham2018, vandesande2021}, denoted as `fast rotators' (or regular rotators) and `slow rotators' by ATLAS$^{\rm{3D}}$ \citep{emsellem2007}. As a first step, we therefore separate slow rotator galaxies from the fast rotators using the cut of \citet{cappellari2016}:
\begin{equation}
    \lambda_{Re} < 0.08 + \epsilon/4 ~~~~\rm{and}~~~~ \epsilon<0.4,
    \label{eqn}
\end{equation}
where $\epsilon$ is the observed ellipticity.
Any galaxy that does not satisfy the criteria of Equation~\ref{eqn} is designated as a fast rotator galaxy.
We plot the distributions of both fast rotators (purple histogram) and slow rotators (red, unfilled histogram) in $\log V/\sigma$ in Figure~\ref{fig:results1} in four bins of stellar mass. At all stellar masses, the distribution of $\log V/\sigma$ comprises a clear peak at $\log V/\sigma\sim0$ (which we expect to be a population of dynamically-cold disks) and a long tail to low $\log V/\sigma$, which becomes more prominent at high stellar masses. 
In particular, for the highest mass bin, the peak at high $\log V/\sigma$ is less well-defined, and a secondary peak of `slow rotators' at low $\log V/\sigma$ becomes prominent.

Modelling the distribution of fast rotators is far from trivial, as it is not clear how many populations are present. The long tail to low $\log V/\sigma$ could be indicative of a single (continuous) distribution, or a bi or trimodal population.
There has been previous work devoted to carefully modelling populations of galaxies in kinematic space \citep[e.g.][]{vandesande2021}. In this work, we take a different (and slightly simpler) approach, and start from the assumption that the distribution of $\log V/\sigma$ for dynamically-cold disk systems is well approximated by a normal distribution \citep[i.e. matching the distribution of spin parameters of dark matter halos in galaxy formation simulations e.g.][]{mo1998}. This assumption allows us to confidently separate dynamically cold systems\footnote{Nomenclature can often open a can of worms. Here, we refer to galaxies that reside at the peak of the $\log V/\sigma$ distribution that are consistent with observations of dynamically-cold, kinematically axisymmetric, rotating systems with the lowest contribution of random motions within $1R_{e}$ as dynamically-cold disks. While it is possible that these galaxies contain multiple components, they are still the most rotationally-dominated systems in the local Universe. From the point of view of traditional photometric bulge-disk decompositions, these galaxies would be considered `pure' disks.} from the dynamically warmer population of fast-rotators that, hereafter, we will denote as `intermediate systems'.

We model the fast rotator population (purple histogram in Figure~\ref{fig:results1}) with a Gaussian mixture model applied to the $\log(V/\sigma)$ distribution. For all stellar mass bins, the best fit was found using two components. We note that we do not necessarily expect the lower $\log(V/\sigma)$ galaxies to be modelled by a Gaussian, but instead use the mixture model to properly model the dynamically-cold disk population that peaks at $\log(V/\sigma)\sim0$ (shown as the navy line in Figure~\ref{fig:results1}). As we are interested in the importance of disks across the star-forming main sequence, it is important to create as clean a population of dynamically-cold disks as possible. For this reason, we define the conservative cut of any galaxy with $\log(V/\sigma)>-1\sigma$ from the mean of the Gaussian that is fit to the higher-$\log(V/\sigma)$ population as being a dynamically-cold disk. The dividing line between the two populations is shown as a grey dashed line in Figure~\ref{fig:results1}, and in order of ascending stellar mass, occurs at $\log V/\sigma=$ -0.14, -0.10, -0.14, and -0.11 respectively.
Any fast rotator with $\log V/\sigma$ lower than this value is designated an intermediate system. The $1\sigma$ cut chosen is an upper limit on the dynamically-cold disk population, but ensures minimal contamination from intermediate systems in the dynamically-cold disk category. Indeed, these cuts represent a $>90\%$ probability of a galaxy being a dynamically-cold disk when the Gaussian mixture model probabilities are examined. While our stringent cut ensures a clean dynamically-cold disk sample, we note that the intermediate systems are likely a heterogeneous population, including galaxies with significant thick disks, dispersion-supported bulges, or other dynamically-warm structures responsible for reducing their $V/\sigma$.
A hard cut also ensures that our work may be easily reproduced by other studies. 

We choose to perform the following kinematic analysis using the $V/\sigma$ parameter rather than $\lambda_{Re}$ as it is approximately log normally distributed. By design, the $\lambda_{Re}$ parameter has an artificial ceiling at $\lambda_{Re}=1$, and while this parameter does include extra information on the relative radial distribution of flux within a galaxy, it cannot be used to model a disky `main sequence'.

As can be seen from Fig~\ref{fig:results1}, for all stellar mass bins, the galaxy population is dominated by fast rotators. As expected, the fraction of slow rotators increases with increasing stellar mass, with the population comprising the most significant fraction of galaxies at $\log M_{\star}[\rm{M}_{\odot}]>11.25$.  Overall, we find that slow rotators are not the dominant galaxy morphology at any stellar mass in line with previous works \citep[e.g.][]{vandesande2017a, guo2020}.
Given the three populations of galaxies: dynamically-cold disks, intermediate systems, and slow rotators, we investigate the prevalence of each as a function of a galaxy's distance from the star-forming main sequence line, $\Delta \rm{MS}$, as defined in \citet{fraser-mckelvie2021}.

\subsection{Kinematic morphology as a function of SFR}
\label{sect_31}
In Figure~\ref{fig:pyramid}, we plot the distribution of $\Delta$MS for dynamically-cold disks (blue, left), intermediate systems (green, centre), and slow rotators (maroon, right) for four bins of stellar mass. The overall galaxy distribution for the MaNGA kinematic sample for each mass bin is shown in grey. There are obvious differences in the star formation activity of each kinematic class: dynamically-cold disks, for example, dominate the star-forming main sequence at $\Delta \rm{MS}=0$ \citep[as shown in][]{wang2020,fraser-mckelvie2021}.
In fact, if we exclude this class of objects, there is no star-forming main sequence when looking at intermediate systems and slow rotators. Of course there are still some star-forming intermediate and slow rotator systems, but the distribution across $\Delta$MS is wider, without a clear peak. Similarly, if we focus on the passive population, it is almost always dominated by intermediate systems, (and not slow rotators) but there is always also a population of passive dynamically-cold disks. The diversity of kinematic morphologies in the passive regions highlight how the bimodality in SFR does not map perfectly into structure as the structure distribution is at least trimodal. We explore the star formation activity of each kinematic class and speculate on possible drivers below.

\subsubsection*{Dynamically-cold disks}
The majority of dynamically-cold disk galaxies are star-forming. We may explain this observation by the expectation that star-forming disk galaxies have kept growing and acquiring angular momentum over the last few billion years, either via accreted cold gas from the intergalactic medium or galactic fountain effects. Cold mode gas accretion is expected to be anisotropic and filamentary in nature, possessing on average high specific angular momentum \citep[e.g.][]{pichon2011,stewart2011,stewart2013}. The addition of high angular momentum cold gas spins up the galaxy, leaving it with greater angular momentum compared to galaxies that have not recently accreted gas. \citep[e.g.][]{white1984, lagos2017, cortese2019}. Simulations show that in most cases, galactic fountains acquire angular momentum via the condensation of high-angular momentum circum-galactic medium (CGM) gas, which proceeds to rain down on the galaxy, further extending its disk \citep[e.g.][]{fraternali2008,marinacci2010,brook2012,defelippis2017,grand2019}. Galactic fountains require stellar feedback, and so the addition of angular momentum via fountain mechanisms may be something of a Catch 22: to gain both angular momentum and cold gas from the CGM, a galaxy requires feedback via stellar winds, and hence must remain star-forming to acquire more fuel for star formation.

\begin{figure*}
\begin{centering}
\includegraphics[trim=2cm 3.5cm 5cm 1cm, clip, width=0.88\textwidth]{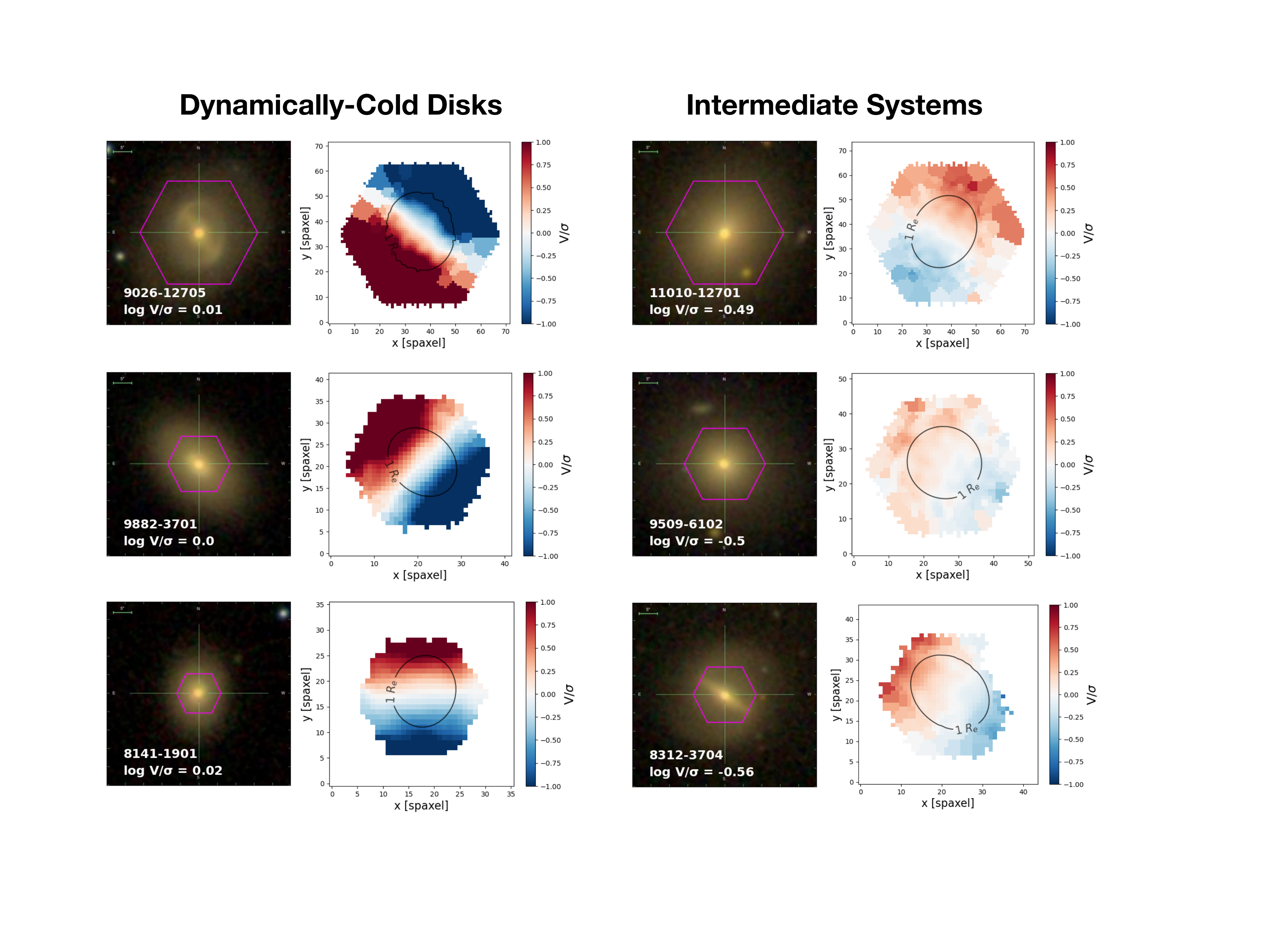} % [left, bottom, right, top]
%Documents/2022/MaNGA_vsig/fig4.key (you could NOT get this working in python or latex!!!) Fig4.py used to create the actual images
\caption{Examples of SDSS $gri$ images (with MaNGA field of view overlaid in magenta) and stellar $V/\sigma$ maps of three passive dynamically-cold disk galaxies (left) and passive intermediate systems (right) in the sample. While the visual morphologies are very similar, there are obvious differences in the $V/\sigma$ maps. \label{fig:passivespirals}}
\end{centering}
\end{figure*}

Passive dynamically-cold disk galaxies are also numerous throughout the mass range studied: they account for 38\%, 33\%, 36\%, and 10\% of all passive galaxies in the mass bins $9.75<\log M_{\star}[\rm{M}_{\odot}]<10.25$, $10.25<\log M_{\star}[\rm{M}_{\odot}]<10.75$, $10.75<\log M_{\star}[\rm{M}_{\odot}]<11.25$, $11.25<\log M_{\star}[\rm{M}_{\odot}]<11.75$, respectively (and we note that these fractions do not change significantly if we reduce the passive cut from -0.5 dex below the star-forming main sequence to -0.9 dex). Examples of SDSS 3-color images of three passive dynamically-cold disk galaxies and their stellar $V/\sigma$ maps are shown in the left columns of Figure~\ref{fig:passivespirals}. 
Combining all mass bins in this work, the total fraction of passive galaxies that possess dynamically-cold disk morphology at $9.75<\log M_{\star}[\rm{M}_{\odot}]<11.75$ is 35\%.

The origin of passive spiral galaxies has been debated in the literature for some time, though the two most likely candidates are group or cluster-related stripping processes \citep[e.g.][]{poggianti1999, goto2003}, and internal secular processes including stellar bars \citep[e.g.][]{masters2010, fraser-mckelvie2018b}. Briefly, environmental processes such as ram-pressure stripping are expected to remove gas from galaxies in a manner gentle enough so as not to disrupt the spiral arm structure. Also possible is the accelerated gas usage as the result of bar funnelling processes acting to use up a galaxy's gas supply before it can be replenished. While the origin of the passive dynamically-cold disks in our sample is beyond the scope of this work, we note that there are certainly mechanisms available to quench a galaxy without significantly disrupting the stellar rotational support of its disk \citep[e.g.][]{cortese2021}. 

We speculate that from a statistical point of view, the passive dynamically-cold disk population may have quenched at later times than the intermediate population: they had enough time to benefit from the extra angular momentum imparted on them via gas accretion or galactic fountain effects, and have not had the chance for interactions or any other processes that reduce spin to transpire. The above scenario is in agreement with the observation that the highest $V/\sigma$ galaxies in the dynamically-cold disk sample are star-forming: the passive disks, whilst still possessing high $V/\sigma$, are slightly less rotation-suppported than the star-forming sample. 

The most common morphology of star-forming galaxies for all stellar masses is dynamically-cold disks. This observation is at odds with photometric works such as \citet{lang2014} and \citet{morselli2017}, who found that the dominant morphology of star-forming galaxies at high stellar masses is bulge+disk systems, which they explained to be the result of compaction-type processes \citep[e.g.][]{tacchella2016}. Indeed, even when a more severe cut (e.g. $\log V/\sigma>0$) is used, we find that dynamically-cold disks remain the dominant class of objects on the star-forming main sequence.

Recent work that has probed the inner regions of galaxies in extreme detail has found that multiple kinematic components exist: nuclear disks are very common, as are nuclear star clusters and inner bars \citep[e.g.][]{gadotti2020}. The excess light above a disk surface brightness profile could be attributed to the superposition of these components, which are still disky in nature \citep[e.g.][]{erwin2021}. Indeed, when careful photometric structural decomposition is carried out, the fraction of highly star-forming bulge-dominated systems decreases, a fact attributed to the limited ability for viable model validation of large samples \citep[e.g.][]{cook2020}.
The presence of kinematically cold structures that contribute to a bulge-like light distribution may explain why we see no evidence of traditional bulge+disk systems in the star-forming region of our sample. 

\subsubsection*{Slow Rotators}
Galaxies may reside in the region of the $\lambda_{Re}-\epsilon$ plane typical of slow rotators for multiple reasons. Most commonly, their stellar orbits are dominated by random motions and the galaxy itself is dispersion-supported \citep[i.e. they are true spheroids, see][]{cappellari2016}. This scenario is most likely to be the case for the majority of galaxies labelled as slow rotators in Figures~\ref{fig:results1} and~\ref{fig:pyramid}. Galaxies can also fall into this category if they possess ordered rotation in multiple directions within 1$R_{e}$: for example, counter-rotating cores, or if they are undergoing or have undergone a significant interaction that has disrupted the stellar kinematics. In these cases, despite possessing a spin parameter consistent with disordered rotation, the galaxy is not a spheroid. For the small numbers of star-forming (and low-mass) slow rotators seen in Figures~\ref{fig:results1} and~\ref{fig:pyramid}, we expect that this is the case. 

The star-forming slow rotators are a small, but interesting population. At high stellar masses, they simply constitute the tail of the passive population, most likely due to the unreliability of our SED-derived SFRs for specific SFRs $<10^{-11.5}$ \citep{salim2018}. This assumption is confirmed by the fact that upon visual inspection, these galaxies appear to be genuinely spheroidal systems. 
We expect this issue for only a small portion of the overall population, and so do not expect these SFRs to make an appreciable difference to the overall galaxy distribution.

Star-forming slow rotators at lower stellar masses are still rare, but they become a very heterogeneous population. Indeed, upon careful visual inspection of the 14 star-forming slow rotators below $\log M_{\star}[\rm{M}_{\odot}]<10.75$, five are interacting, %7993-1902, 8440-6104, 8997-12702, 8997-12705, 12514-3702
one is a bona fide slow rotator with very passive spectra, %11744-12702
one is a kinematically-decoupled core galaxy, %8615-1902
five are face-on disk galaxies, whose line-of-sight velocity contribution is too low to classify them as fast rotators, %12700-1902, 8440-3703, 8946-37001, 9869-9012, 9193-3704
and two have $R_{e}$ that are very close to that of the PSF of MaNGA. 
%(2.86$^{\prime\prime}$ and 2.66$^{\prime\prime}$ respectively). %8245-3701, 12487-1902 )

Overall, the fraction of slow rotators begins to become appreciable only in the two highest stellar mass bins, and nearly all of which are passive. We expect that this is the population dominated by true spheroidal galaxies: those whose stars are dominated by random motions in an equilibrium configuration. The incidence of true spheroids is relatively rare: the fraction of slow rotators in the two highest mass bins is 10\% and 27\%. We expect the high-mass slow rotator population to be relatively homogeneous: long-quenched galaxies who have experienced destructive gravitational interactions that have made these systems almost entirely dispersion supported.

Across the entire mass range probed, the total slow rotator fraction is 6\%. This value is low compared to earlier works: for example, 14\% in ATLAS$^{\rm{3D}}$ \citep{emsellem2011}, and 15\% in both SAMI early data \citep{fogarty2014} and Coma \citep{houghton2013}. While the stellar mass range probed is different in every case, most of these earlier works consisted of small samples and/or specific morphologies and environments. A lack of a representative sample, coupled with uncorrected beam-smearing effects means the slow rotator fraction is possibly over-reported compared to this current work.
\citet{vandesande2017a} found that 8.6\% of galaxies from the SAMI galaxy survey data release 2 (DR2) were slow rotators, which was revised to 9.9\% for the full SAMI DR3 sample \citep{vandesande2021a}. Neither of these works employed a volume correction to the SAMI data, and along with differences in the stellar mass range probed, we expect this to be the dominant factor for the differences seen (see Section~\ref{mass_budget} for a further comparison to volume-corrected SAMI data). 

\subsubsection*{Intermediate systems}
The intermediate class of galaxies is arguably the most interesting, as despite being primarily passive, we expect this population to be heterogeneous in their structural makeup and evolutionary histories. The intermediate population comprises galaxies within the tail of the $\log V/\sigma$ histograms seen in Figure~\ref{fig:results1}. These galaxies possess lower $V/\sigma$ than those designated as dynamically-cold disks, though not so low as to be classified as a slow rotator: some degree of rotational support is still present, and these galaxies would still be classified as fast rotators using the ATLAS$^{\rm{3D}}$ nomenclature. Examples of three passive intermediate systems are shown in the right columns of Figure~\ref{fig:passivespirals}. Visually, they are very similar to the optical images of the dynamically-cold disks, though show clear differences in their $V/\sigma$ maps.

There are multiple reasons why a passive galaxy may not possess the maximum spin expected for its stellar mass.
One option is that a galaxy was star-forming and rotation-dominated in the past, but either coincidentally or separately, grew dispersion-supported structure which had a net negative contribution to its integrated spin measure. Such structures include classical bulges (possibly built up by interactions), or a dynamically hot thick disk. This idea is supported by photometric observations of higher bulge-to-total ratios in passive galaxies \citep[e.g.][]{baldry2004,bundy2010,cheung2012, bluck2014,morselli2017, popesso2019}, and a link between bulge prominence and quenching in IFS studies \citep{brownson2022}. The growth of both classical bulges and thick disks has been attributed to mergers \citep[e.g.][]{quinn1993,walker1996,aguerri2001, read2008,villalobos2008}, or gas-related compaction events \citep[e.g.][]{tacchella2016}. It is possible that the intermediate systems have had more active merger histories than the dynamically-cold disks. Given that the vast majority of intermediate systems are not star-forming, if the structure growth hypothesis is correct, we can infer that the build-up of such a dispersion-supported component must come either at the same time or later than quenching: we find no evidence of the growth of classical bulges or other dispersion-supported structure prior to galaxy quenching \citep[as in][]{fraser-mckelvie2021, cortese2022}.

Another option for lower-than-expected spin could simply be that the galaxy quenched early and did not have the opportunity to accrete extra high-angular momentum gas from its surroundings. Rather than thinking of the galaxy having `lost' angular momentum via interactions or the growth of dispersion-supported structure, it just never had the opportunity to accrete additional gas and spin up to the levels of star-forming dynamically-cold disks \citep[e.g.][]{pedrosa2015,lagos2017, el-badry2018,cortese2019}.

One final explanation for the intermediate class is that of observational biases. Disk fading, for example, is a process by which a galaxy's disk ceases star formation and dims \citep[e.g.][]{croom2021a}. Such a process results in a galaxy whose light profile is more centrally concentrated, affecting observational parameters including the bulge-to-total ratio and the effective radius, despite there being no structural change to the galaxy. The consequence of disk fading for integrated light-weighted kinematic parameters including $V/\sigma$ and $\lambda_{Re}$ is the increased contribution of central regions to the overall spin measurement, often resulting in lower spin measurements. This scenario highlights the limitations of our current galactic stellar kinematic indicators: without mass-weighted kinematic measures it is impossible to be sure that a lower spin measurement corresponds to a genuine structural difference within a galaxy, or just the effect of a more concentrated radial light profile of a galaxy. 

Given the possibility of a contribution from disk fading, it is entirely feasible that a significant portion of passive intermediate systems from this work are in fact faded dynamically-cold disks, placing a lower limit on the disk fraction in the local Universe. As long as our kinematic indicators are limited to light-weighted quantities, we will not be able to differentiate between the processes causing a galaxy's kinematic measures to place it in the intermediate category. Ideally, mass-weighted kinematic quantities will be the way forward in this respect. 

\begin{figure}%/Documents/2022/MaNGA_vsig/integrated_mass_fractions.py
\gridline{\fig{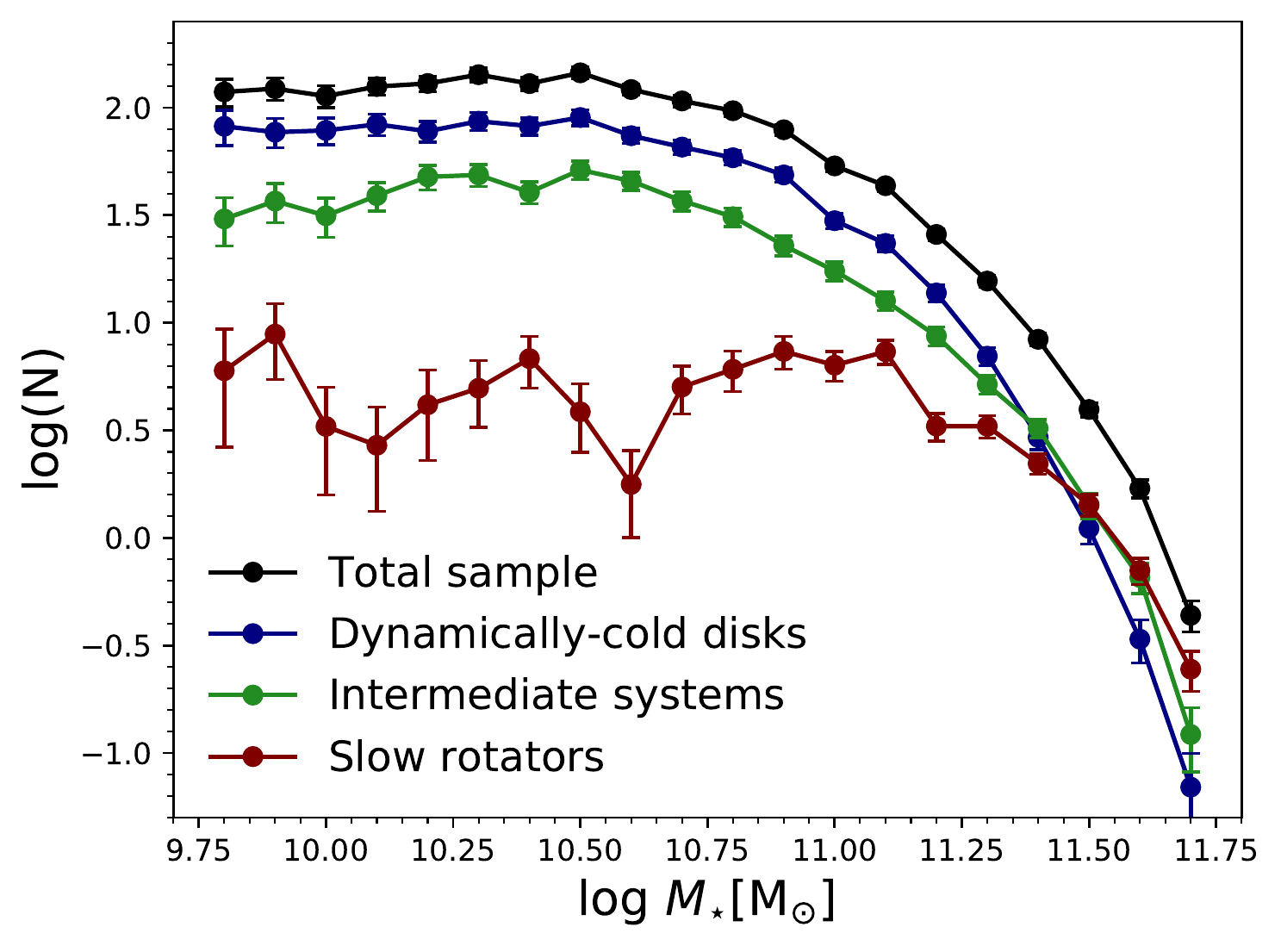}{0.4\textwidth}{}}
\gridline{\fig{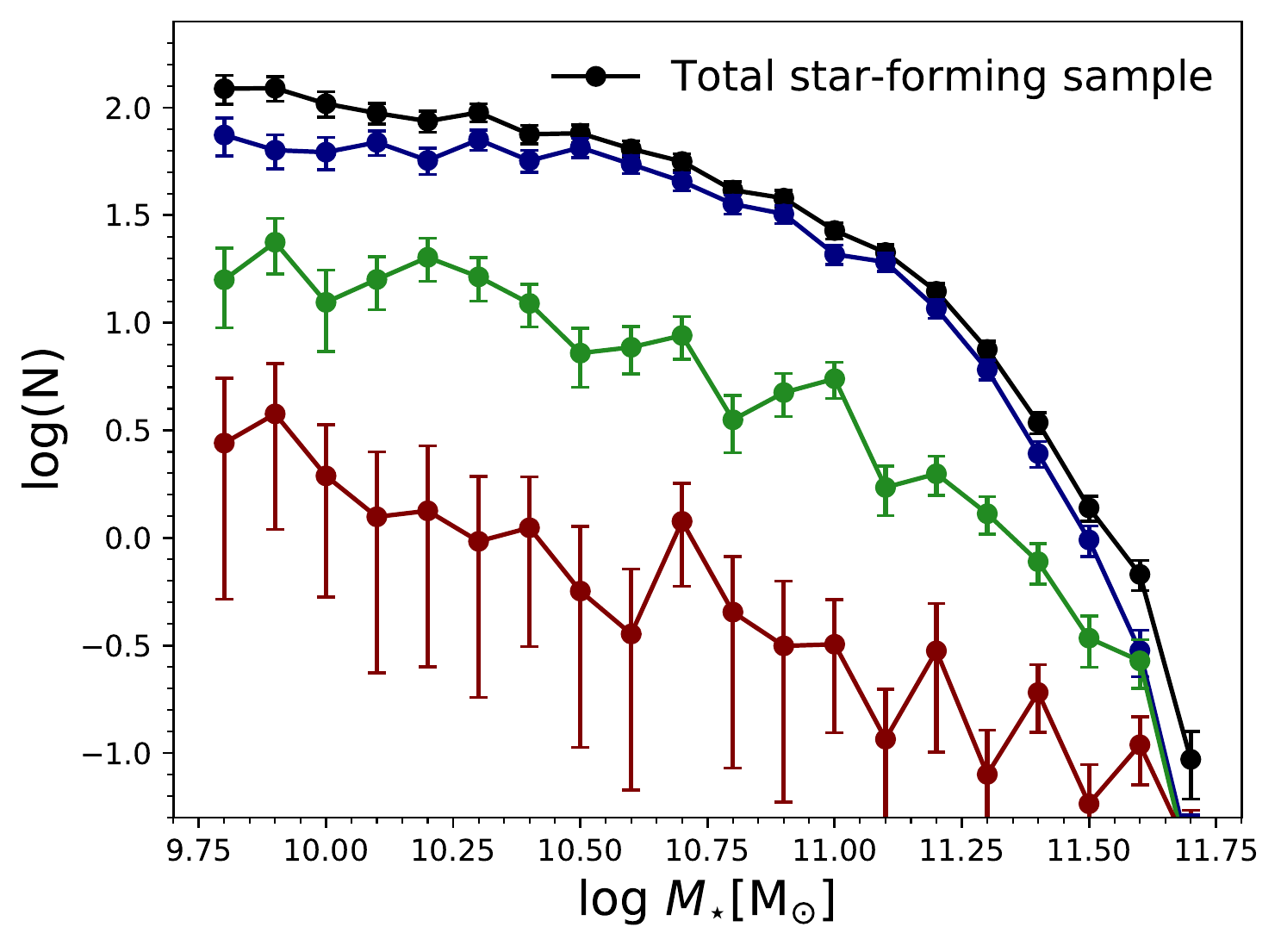}{0.4\textwidth}{}
          }
\gridline{\fig{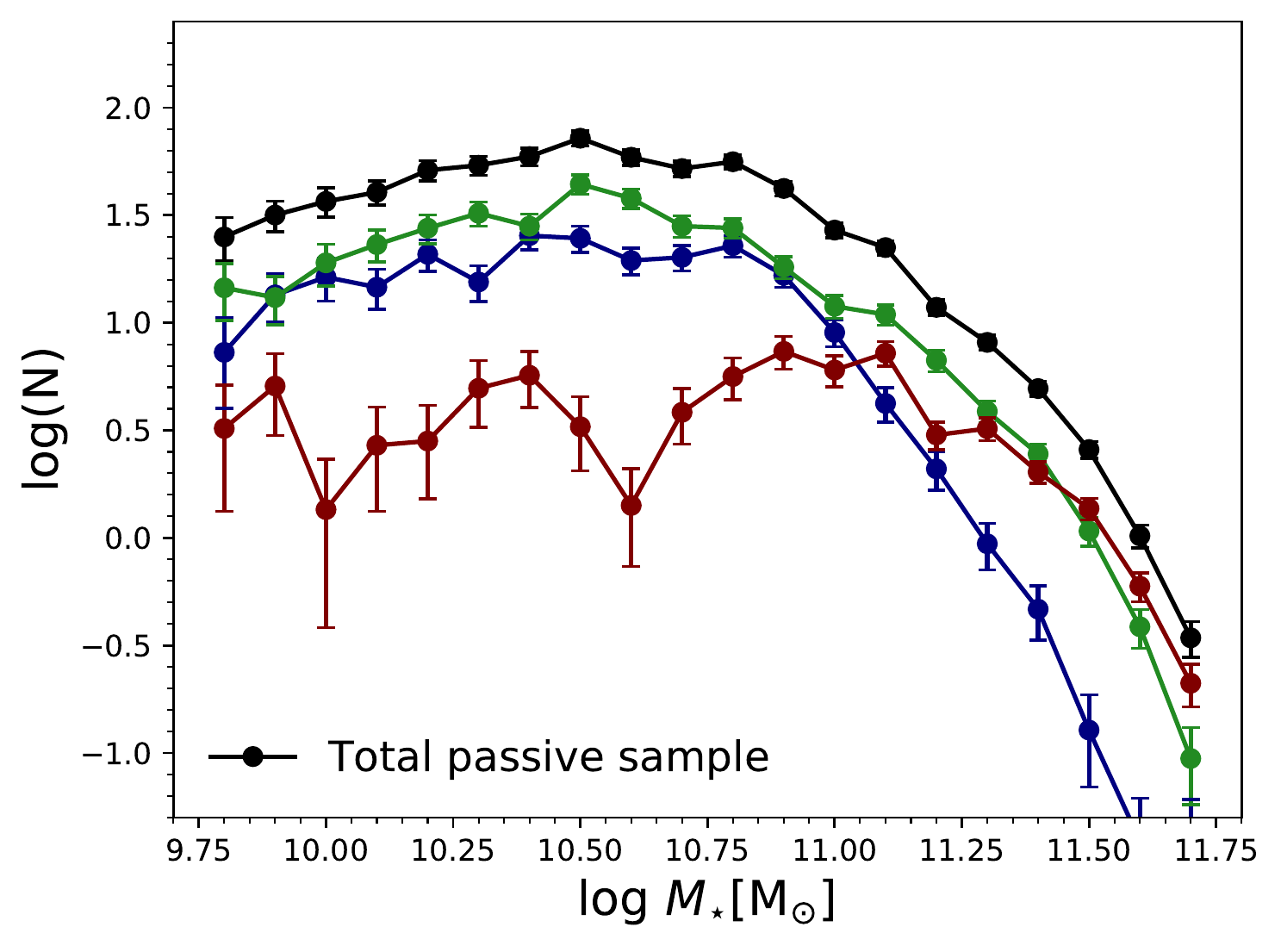}{0.4\textwidth}{}
          }
\caption{Mass distributions of the the total volume-weighted MaNGA kinematic sample (black), and the contribution from dynamically-cold disks (blue), intermediate systems (green) and slow rotators (red). The full sample is shown in the top figure, and in the centre shows the mass distribution for star-forming galaxies only ($\Delta$MS$<-0.5$), while the bottom panel shows the passive population. Dynamically-cold disks dominate the mass budget for star forming galaxies, but intermediate systems dominate in the passive sample. Slow rotators account for an appreciable amount of mass only for passive galaxies with $\log M_{\star}[\rm{M}_{\odot}]>11$. \label{fig:num_mass_fracts}}
\end{figure}

\subsection{Mass budget}
\label{mass_budget}
We quantify the fraction of stellar mass locked up in each kinematic type as a function of stellar mass for the entire sample in Figure~\ref{fig:num_mass_fracts} (top). The total sample is shown as a black line, dynamically-cold disks in blue, intermediate systems in green, and slow rotators in maroon. 
These mass distributions show the relative contribution of each morphological type as a function of stellar mass. 
For the overall population, the dominant contribution of dynamically-cold disks to the mass budget is apparent at all but the highest stellar masses. Intermediate systems also contribute at all stellar masses, but slow rotators only make an appreciable contribution at high stellar masses, above $\log M_{\star}[\rm{M}_{\odot}]=11.25$. Overall, slow rotators account for just 6\% of all galaxies and 10\% of the total mass budget between $9.75< \log M_{\star}[\rm{M}_{\odot}]<11.75$. Number and mass fractions for each kinematic type are listed in Table~\ref{table1}.

\citet{guo2020} quantified the mass budget of different kinematic classes for 1896 galaxies from the SAMI galaxy survey. Slow rotators accounted for 14\% of the mass budget of their total sample, compared to just 6\% when the sample in this work is restricted to their mass limit of $\log M_{\star}[M_{\odot}]<11$. We expect the discrepancy comes from the PSF correction applied in this work, which acts to increase the $\lambda_{Re}$ of galaxies in seeing-dominated data. Indeed, when \citet{guo2020} apply the seeing correction of \citet{graham2018} to their data, their slow rotator fraction drops to 8\%.

As discussed in Section~\ref{sect_31} and shown in early ATLAS$^{\rm{3D}}$ results \citep{emsellem2011}, slow rotators contribute only a small amount to the overall number counts and mass budget of the local Universe. That the fraction of slow rotators increases with stellar mass has been shown by \citet{cappellari2013} (ATLAS$^{\rm{3D}}$), \citet{veale2017} (MASSIVE+ATLAS$^{\rm{3D}}$), \citet{vandesande2017a} (SAMI), and \citet{graham2018} (MaNGA). These IFS results are in stark contrast to photometric results, which place spheroids as the dominant morphology of the local galaxy population. \citet{brennan2015} for example, found spheroidal morphology within GAMA at $z=0$ to be 90\% at $11<\log M_{\star}[\rm{M}_{\odot}]<11.5$. Through an analysis of mass functions of disky and spheroidal galaxies also within GAMA, \citet{moffett2016} revised the elliptical fraction to 36\% for all stellar masses. There is a clear disconnect between photometric morphologies and kinematic: more galaxies are allocated to the spheroid class when classified visually. The knock-on effect of this discrepancy is the relative (un)importance of disruptive mergers in the local Universe. IFS survey results would appear to downplay their effects, given that the majority of galaxies do not seem to host the dispersion-supported structures that they are required to build.
Our results lend weight to the importance of considering stellar orbits when classifying true spheroids, and confirm that only at the highest stellar masses do slow rotator galaxies make a significant contribution to the overall galaxy number counts and mass budget of the local Universe. 

In order to further investigate the contribution of each kinematic class as a function of galaxy SFR, we separate the population into star-forming ($\Delta\rm{MS}>-0.5$ dex) and passive ($\Delta\rm{MS}<-0.5$ dex) in Figure~\ref{fig:num_mass_fracts} centre and bottom, respectively (though we note that the following results do not change dramatically when $\Delta\rm{MS}=-0.9$ is used as the dividing line). We see that star-forming systems are dominated by dynamically-cold disk morphology at all stellar masses, whilst slow rotators are scarce. 
When the passive population is considered, it is the intermediate systems that dominate for stellar masses below $\log M_{\star}[\rm{M}_{\odot}]<11.5$. At $\log M_{\star}[\rm{M}_{\odot}]\sim11$, slow rotators begin to dominate in number above dynamically-cold disks, and by $\log M_{\star}[\rm{M}_{\odot}]\sim11.5$ are the most frequent galaxy type. There is a pronounced distinction between the star-forming and passive populations: while the star-forming population is dominated by dynamically-cold disk galaxies, the passive population is split: dynamically-cold disks and intermediate systems dominate below $\log M_{\star}[\rm{M}_{\odot}]<11.5$, but slow rotators are the dominant morphology above this mass. Lastly, we note that below the main sequence, dynamically-cold disks are much more frequent in number (35\% c.f. 10\%) and in total stellar mass budget (30\% c.f. 18\%) than slow rotators. Our results imply that in the stellar mass range investigated here, passive systems are more likely to be dynamically-cold disks than spheroids.

\section{Summary \& Conclusions}
\label{Sect4}
We have performed a kinematic census of galaxies in the local Universe and linked their degree of rotational support to their current star formation activity.
We separate galaxies into three kinematic classes firstly by dividing fast rotator galaxies from slow rotators with a cut in spin parameter $\lambda_{Re}$, and then modelling the fast rotators as a two-component distribution, the dynamically-cold disk population of which may be modelled as a Gaussian in $\log V/\sigma$ space. The tail of this distribution is designated as intermediate systems. The three classes, dynamically-cold disks, intermediate systems, and slow rotators, comprise galaxies of varying rotation-to-dispersion support. We find that:

\begin{itemize}
\item \textbf{Dynamically-cold disks dominate the star-forming main sequence and the overall local Universe mass budget.} We expect that star-forming galaxies were able to acquire additional angular momentum from cold gas accretion or galactic fountain effects at late times, resulting in dynamically cold disks. 

\item \textbf{There is a significant population of passive dynamically-cold disks.} 35\% of all passive galaxies are dynamically-cold disks, meaning that quenching without disk heating or morphological transformation is a viable (and common) pathway in the local Universe. We speculate that at least a fraction of this population has quenched recently, having already benefited from angular momentum deposition via early-times gas accretion. 

\item \textbf{Slow rotators are almost always passive and only contribute appreciably to the total stellar mass budget at $\log M_{\star}[\rm{M}_{\odot}]>11.5$.} IFS surveys attribute a far lower fraction of total stellar mass to spheroidal systems than photometric bulge+disk decompositions. The small fraction of truly dispersion-supported systems, combined with the significant fraction of passive dynamically-cold disks, suggests that ellipticals are not the end point of galaxy evolution for the majority of the local Universe population. 

\item \textbf{Intermediate systems are primarily found below the star-forming main sequence, but likely constitute a heterogeneous population}, including bona fide bulge+disk systems, faded disks, and those that quenched long enough ago that they didn't have access to the additional angular momentum imparted on them by cold mode accretion and/or galactic fountain effects. 

\item \textbf{The distribution of kinematic morphologies is not 1:1 mapped to SFR:} while the majority of star-forming galaxies are dynamically-cold disks, there is a mix of disky, intermediate, and slow rotator morphologies within the passive population. 
\end{itemize}

Our work also highlights the need to move beyond light-weighted spin measures to understand the link between kinematic structure and star formation activity. Greater insight into the consequence of kinematic structural changes to a galaxy could be gained by replicating this work on star-forming galaxies at higher redshift.

Our results paint a picture of a local Universe dominated by disky galaxies. True spheroids are rare and only dominate the mass budget at very high stellar masses. The diversity of kinematic structure present in the passive population reveals a myriad of evolutionary mechanisms.

\appendix
\section{Volume weighting}
\label{appendix1}
\begin{figure*} %calc_volume_weights.py
\plotone{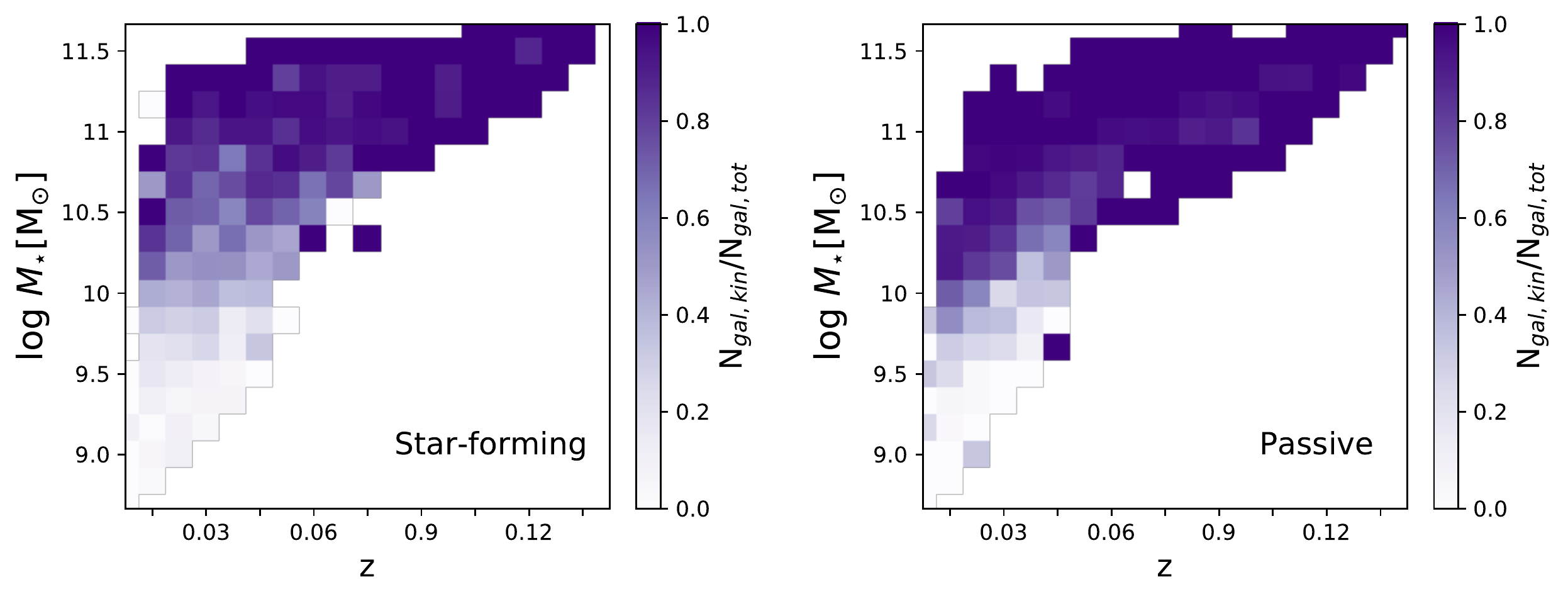}
\caption{Completeness fractions for star-forming (left) and passive (right) samples as a function of redshift and stellar mass. These fractions are used to calculate amended volume weights for the incomplete MaNGA sample used in this work.  \label{fig:appendix1}}
\end{figure*}

In Appendix~\ref{appendix1}, we provide further information on the volume weighting employed in this work. In Figure~\ref{fig:appendix1} we plot the completeness, $N_{gal,kin}/N_{gal,tot}$, of the kinematic sample in bins of redshift and stellar mass. A completeness of 1 indicates that kinematic measures were successfully computed for all galaxies in the MaNGA sample for a given bin. As expected, higher-mass galaxies were more likely to have kinematic measures calculated, whereas low-mass galaxies frequently suffered from low signal-to-noise ratio and/or masking due to low $\sigma_{\star}$. Given that \citet{sanchez2019} found that completeness can vary as a function of galaxy SFR, we split the sample into two: star-forming galaxies Fig~\ref{fig:appendix1} (left), and passive galaxies (right). There are small differences in the completeness fraction at a given redshift and stellar mass between the two.

\begin{figure*}
\plotone{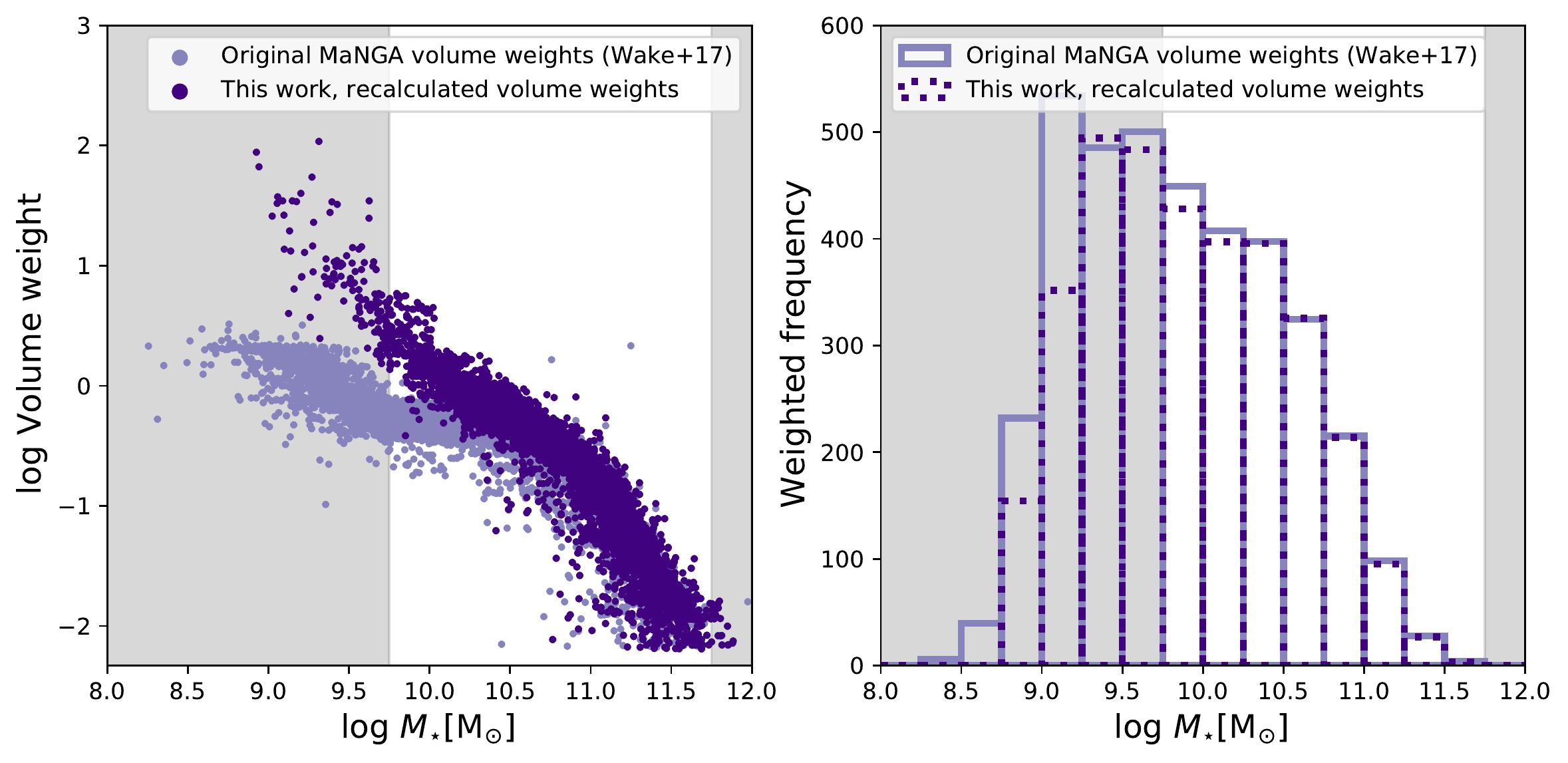} %calc_volume_weights.py
\caption{Volume weighting diagnostics. On the left, we plot the volume weights assigned to each galaxy in our kinematic sample by \citet{wake2017} (mauve), and those recalculated based on completeness as a function of stellar mass and redshift (dark purple). The difference in volume correction is most apparent where we are least complete: at low stellar masses. In the right panel we plot a histogram of the MaNGA volume-corrected stellar mass function (mauve), and that made using the volume corrections calculated for this work (dark purple). We show that our technique for volume weighting recovers the mass function well for stellar masses of $\log M_{\star}[\rm{M}_{\odot}]>9.75$. In both panels, the shaded grey region indicates the stellar mass range excluded from this work. \label{fig:appendix2}}
\end{figure*}

After adjusting the supplied MaNGA volume weights by the derived completeness fractions, we plot the difference between the two volume weights in Figure~\ref{fig:appendix2} (left). Given that the fraction of galaxies with kinematic measures drops significantly at low stellar masses, the recalculated weights differ the most from the supplied weights here. In Figure~\ref{fig:appendix2} (right), we plot a histogram of the original MaNGA weighted (mauve) and reweighted (dark purple) stellar masses. The reweighted histogram matches that of the original MaNGA volume weights well above stellar masses of $\log M_{\star}[\rm{M}_{\odot}]=9.75$, and for this reason, we perform our analysis only for stellar masses higher than this value.

\begin{table*}
\begin{center}
\begin{tabular}{ |l|c|c|c|c|c|c| } 
 \hline
  &\multicolumn{2}{|c|}{\textsc{Dynamically-cold disks}} & \multicolumn{2}{|c|}{\textsc{Intermediate Systems}} & \multicolumn{2}{|c|}{\textsc{Slow rotators}}  \\
  \hline
 & \textbf{Number} & \textbf{Mass} & \textbf{Number} & \textbf{Mass}& \textbf{Number} & \textbf{Mass}\\ 
 \textbf{Mass range}   & \textbf{fraction} & \textbf{fraction} & \textbf{fraction} & \textbf{fraction}& \textbf{fraction} & \textbf{fraction}\\ 
 \hline
 \hline 
 \multicolumn{7}{|l|}{\textbf{Total sample}}\\
\hline
 $9.75 < \log M_{\star}[\rm{M}_{\odot}]<10.25$ &0.63$\pm0.04$ & 0.62$\pm0.04$ &0.33$\pm0.02$ & 0.34$\pm0.02$ & 0.04$\pm0.01$ & 0.04$\pm0.01$ \\ 
 $10.25 < \log M_{\star}[\rm{M}_{\odot}]<10.75$ &0.57$\pm0.02$ & 0.57$\pm0.02$ & 0.39$\pm0.02$ &0.39$\pm0.02$ & 0.04$\pm0.01$ & 0.04$\pm0.01$ \\ 
 $10.75 < \log M_{\star}[\rm{M}_{\odot}]<11.25$ &0.58$\pm0.02$ & 0.57$\pm0.02$ &0.32$\pm0.01$ & 0.32$\pm0.01$ &0.10$\pm0.01$ & 0.11$\pm0.01$\\
 $11.25 < \log M_{\star}[\rm{M}_{\odot}]<11.75$ &0.38$\pm0.03$ & 0.36$\pm0.02$& 0.35$\pm0.02$ & 0.35$\pm0.02$ &0.27$\pm0.02$ & 0.29$\pm0.02$\\
 \hline
    Total sample & 0.59$\pm0.02$ & 0.55$\pm0.02$ & 0.35$\pm0.01$ & 0.35$\pm0.01$ & $(6\pm0.4)\times10^{-2}$ & 0.10$\pm0.01$ \\
 \hline
\hline
\multicolumn{7}{|l|}{\textbf{Star-forming galaxies}}\\
\hline
 $9.75 < \log M_{\star}[\rm{M}_{\odot}]<10.25$ & 0.77$\pm0.06$ & 0.77$\pm0.06$ & 0.21$\pm0.02$ & 0.21$\pm0.02$  & $(2\pm0.2)\times10^{-2}$ & $(2\pm0.2)\times10^{-2}$ \\
 $10.25 < \log M_{\star}[\rm{M}_{\odot}]<10.75$ & 0.81$\pm0.04$ & 0.82$\pm0.04$ & 0.18$\pm0.01$ & 0.17$\pm0.01$ & $(1\pm0.04)\times10^{-2}$ & $(1\pm0.05)\times10^{-2}$ \\
 $10.75 < \log M_{\star}[\rm{M}_{\odot}]<11.25$ & 0.87$\pm0.04$ & 0.86$\pm0.04$ & 0.12$\pm0.01$ & 0.13$\pm0.01$ & $(1\pm0.04)\times10^{-2}$ & $(1\pm0.04)\times10^{-2}$ \\
 $11.25 < \log M_{\star}[\rm{M}_{\odot}]<11.75$ & 0.76$\pm0.06$ & 0.73$\pm0.06$ & 0.21$\pm0.02$ & 0.22$\pm0.02$ & $(3\pm0.3)\times10^{-2}$ & $(5\pm0.4)\times10^{-2}$ \\
\hline
Total sample & 0.79$\pm0.03$ & 0.82$\pm0.03$ & 0.19$\pm0.007$ & 0.16$\pm0.006$ & $(2\pm0.07)\times10^{-2}$ & $(2\pm0.06)\times10^{-2}$ \\
 \hline
\hline
\multicolumn{7}{|l|}{\textbf{Passive galaxies}}\\
\hline
 $9.75 < \log M_{\star}[\rm{M}_{\odot}]<10.25$ & 0.38$\pm0.04$ & 0.39$\pm0.04$ & 0.54$\pm0.05$ & 0.54$\pm0.05$ & 0.08$\pm0.01$ & 0.07$\pm0.01$ \\
 $10.25 < \log M_{\star}[\rm{M}_{\odot}]<10.75$ & 0.33$\pm0.02$ & 0.33$\pm0.02$ & 0.60$\pm0.04$ & 0.61$\pm0.04$ & $(7\pm0.4)\times10^{-2}$ & $(6\pm0.4)\times10^{-2}$ \\
 $10.75 < \log M_{\star}[\rm{M}_{\odot}]<11.25$ & 0.36$\pm0.02$ & 0.33$\pm0.02$ & 0.46$\pm0.03$ & 0.46$\pm0.03$ & 0.18$\pm0.01$ & 0.21$\pm0.01$ \\
 $11.25 < \log M_{\star}[\rm{M}_{\odot}]<11.75$ & 0.10$\pm0.02$ & 0.09$\pm0.01$ & 0.46$\pm0.08$ & 0.45$\pm0.08$ & 0.44$\pm0.07$ & 0.46$\pm0.08$ \\
\hline
Total sample & 0.35$\pm0.02$ & 0.30$\pm0.01$ & 0.55$\pm0.02$ & 0.52$\pm0.02$ & $0.10\pm0.005$ & 0.18$\pm0.01$ \\
\hline
\end{tabular}
\end{center}
\caption{Mass and number fraction of each kinematic type for each mass bin and the total sample, and split into the star-forming and passive populations by a cut at $\Delta$MS=-0.5.  \label{table1}}
\end{table*}

%% IMPORTANT! The old "\acknowledgment" command has be depreciated. It was
%% not robust enough to handle our new dual anonymous review requirements and
%% thus been replaced with the acknowledgment environment. If you try to 
%% compile with \acknowledgment you will get an error print to the screen
%% and in the compiled pdf.
\begin{acknowledgments}
The authors wish to thank David Wake, Kyle Westfall, and Jesse van de Sande for useful conversations related to this work, and the anonymous referee for constructive comments that improved the clarity and quality of this work.
 
This research was conducted by the Australian Research Council Centre of Excellence for All Sky Astrophysics in 3 Dimensions (ASTRO 3D), through project number CE170100013. 
LC acknowledges support from the Australian Research Council Discovery Project and Future Fellowship funding schemes (DP210100337 and FT180100066)

Funding for the Sloan Digital Sky 
Survey IV has been provided by the 
Alfred P. Sloan Foundation, the U.S. 
Department of Energy Office of 
Science, and the Participating 
Institutions. SDSS-IV acknowledges support and 
resources from the Center for High 
Performance Computing  at the 
University of Utah. The SDSS 
website is www.sdss.org.

SDSS-IV is managed by the 
Astrophysical Research Consortium 
for the Participating Institutions 
of the SDSS Collaboration including 
the Brazilian Participation Group, 
the Carnegie Institution for Science, 
Carnegie Mellon University, Center for 
Astrophysics | Harvard \& 
Smithsonian, the Chilean Participation 
Group, the French Participation Group, 
Instituto de Astrof\'isica de 
Canarias, The Johns Hopkins 
University, Kavli Institute for the 
Physics and Mathematics of the 
Universe (IPMU) / University of 
Tokyo, the Korean Participation Group, 
Lawrence Berkeley National Laboratory, 
Leibniz Institut f\"ur Astrophysik 
Potsdam (AIP),  Max-Planck-Institut 
f\"ur Astronomie (MPIA Heidelberg), 
Max-Planck-Institut f\"ur 
Astrophysik (MPA Garching), 
Max-Planck-Institut f\"ur 
Extraterrestrische Physik (MPE), 
National Astronomical Observatories of 
China, New Mexico State University, 
New York University, University of 
Notre Dame, Observat\'ario 
Nacional / MCTI, The Ohio State 
University, Pennsylvania State 
University, Shanghai 
Astronomical Observatory, United 
Kingdom Participation Group, 
Universidad Nacional Aut\'onoma 
de M\'exico, University of Arizona, 
University of Colorado Boulder, 
University of Oxford, University of 
Portsmouth, University of Utah, 
University of Virginia, University 
of Washington, University of 
Wisconsin, Vanderbilt University, 
and Yale University.
\end{acknowledgments}

\bibliography{MaNGA_kins}{}
\bibliographystyle{aasjournal}

%% This command is needed to show the entire author+affiliation list when
%% the collaboration and author truncation commands are used.  It has to
%% go at the end of the manuscript.
%\allauthors

%% Include this line if you are using the \added, \replaced, \deleted
%% commands to see a summary list of all changes at the end of the article.
%\listofchanges

\end{document}